\documentclass[10pt]{iopart}

\usepackage{iopams}
\usepackage{graphicx} % Include figure files
\usepackage{dcolumn}% Align table columns on decimal point
\usepackage{times}
\usepackage{textcomp}
\usepackage{titlesec}
\usepackage{float}
\usepackage{color,xcolor}
\usepackage{graphicx,times} 
\usepackage{cuted} \stripsep -3pt plus 3pt minus 2pt
\usepackage{epstopdf}
\usepackage{graphicx}
\usepackage{subfigure}
\usepackage{makecell}
\usepackage{multirow}
\renewcommand{\figurename}{Figure}
\usepackage[colorlinks,linkcolor=blue, urlcolor=blue, anchorcolor=blue, citecolor=blue]{hyperref}
\usepackage{longtable,booktabs}
\usepackage{fouriernc}
\usepackage{epsfig,graphics,graphicx}
\usepackage{bm}
\usepackage{color}
\usepackage{flushend}
\usepackage{ulem}
\usepackage[sort&compress,square,numbers]{natbib}
\usepackage[final]{pdfpages}

\begin{document}
\title{Fusion Reactivities with Drift bi-Maxwellian Ion Velocity Distributions}

\author{
Huasheng Xie$^{1,2}$, 
Muzhi Tan$^{1,2}$, 
Di Luo$^{1,2}$, 
Zhi Li$^{1,2}$, 
and Bing Liu$^{1,2}$
}

\address{

$^1$ Hebei Key Laboratory of Compact Fusion, Langfang 065001, People’s Republic of China

$^2$ ENN Science and Technology Development Co., Ltd., Langfang 065001, People’s Republic of China

}
\eads{\mailto{huashengxie@gmail.com, xiehuasheng@enn.cn}}

\begin{indented}
\item[\today]
\end{indented}

\begin{abstract}
{The calculation of fusion reactivity involves a complex six-dimensional integral that takes into account the fusion cross section and velocity distributions of two reactants. However, a more simplified one-dimensional integral form can be useful in certain cases, such as for studying fusion yield or diagnosing ion energy spectra. This simpler form has been derived in a few special cases, such as for a combination of two Maxwellian distributions, a beam-Maxwellian combination, and a beam-target combination, and can greatly reduce computational costs. In this study, it is shown that the reactivity for two drift bi-Maxwellian reactants with different drift velocities, temperatures, and anisotropies can also be reduced to a one-dimensional form, unifying existing derivations into a single expression. This result is used to investigate the potential enhancement of fusion reactivity due to the combination of beam and temperature anisotropies. For relevant parameters in fusion energy, the enhancement factor can be larger than 20\%, which is particularly significant for proton-boron (p-B11) fusion, as this factor can have a significant impact on the Lawson fusion gain criteria.}
\end{abstract}
\submitto{\PPCF}
\maketitle
\ioptwocol

\section{Introduction}\label{sec:intro}

The counting of fusion reactions per unit volume and per unit time is\cite{Atzeni2004,Clayton1983}:
\begin{equation}
R_{12}=\frac{n_1n_2}{1+\delta_{12}}\langle\sigma v\rangle,
\end{equation}
where $n_1$ and $n_2$ are the number densities of the two reactants, respectively. The term $\delta_{12}$ is equal to 0 for different reactants and equal to 1 for the same reactants, in order to prevent double counting of the reaction.

Here, $\sigma=\sigma(E)$ or $\sigma=\sigma(v)$ represents the fusion cross section, with $E$ being the energy in the center-of-mass frame, defined as
\begin{equation}
E=\frac{1}{2}m_rv^2,~~v=|{\bm v}|=|{\bm v}_1-{\bm v}_2|,~~m_r=\frac{m_1m_2}{m_1+m_2},
\end{equation}
where $m_1$ and $m_2$ denote the mass of the two reactants, and $m_r$ represents the { reduced mass of the system}.
The fusion reactivity $\langle\sigma v\rangle$ is calculated as the integral of the fusion cross section and the velocity distribution functions of the reactants
\begin{equation}\label{eq:sgmv}
\langle\sigma v\rangle=\int\int d{\bm v}_1d{\bm v}_2\sigma(|{\bm v}_1-{\bm v}_2|)|{\bm v}_1-{\bm v}_2|f_1({\bm v}_1)f_2({\bm v}_2),
\end{equation}
where $f_1,f_2$ are normalized velocity distribution functions of the two ions, i.e., $\int f_{j}({\bm v}_j)d{\bm v}_j=1$ with $j=1,2$, and $d{\bm v}_j=dv_{xj}dv_{yj}dv_{zj}$. Eq.(\ref{eq:sgmv}) is crucial not only for calculating the fusion yield in laboratory or stellar plasmas\cite{Atzeni2004,Clayton1983}, but also for diagnostics\cite{Appelbe2011,Li2022a} by providing information about the spectrum of the distributions $f_{1,2}$. However, calculating $\langle\sigma v\rangle$ is challenging, as it requires a six-dimensional (6D) integral of the velocity, which is usually calculated numerically using high-dimensional integral methods such as Monte-Carlo methods\cite{Lepage1978} or orthogonal polynomials expansion methods\cite{Cordey1978}. If the integral can be reduced to a one-dimensional (1D) or two-dimensional (2D) form, the computation cost would be significantly reduced, and it would also provide analytical insights into the effect of the distributions on reactivity.

Fortunately, in Maxwellian-Maxwellian \cite{Atzeni2004,Clayton1983,Nevins2000} or beam-target\cite{Miley1974,Miley1975,Morse2018} distributions of $f_{1,2}$, the one-dimensional form of $\langle\sigma v\rangle$ can be easily obtained. However, for more general distributions, a simple form of $\langle\sigma v\rangle$ is often not available \cite{Slaughter1983}. Non-Maxwellian fusion reactions are common \cite{Harvey1986,Nakamura2006,Wolle1999}. Recently, there have been efforts to address this issue, such as Nath et al \cite{Nath2013} reducing the integral to three dimensions for drift tri-Maxwellian reactants, Ou et al \cite{Ou2015} studying some beam and target cases, and Kolmes et al \cite{Kolmes2021} (as well as Li et al \cite{Li2022}) providing a one-dimensional form for bi-Maxwellian reactants\cite{Kalra1988} with the use of the error function. For plasma physics studies, one of the most widely used distributions is the drift bi-Maxwellian distribution, which is commonly used in studies of plasma waves and instabilities \cite{Stix1992,Xie2019} and can also be reduced to the isotropic and thermal Maxwellian case. This distribution is also widely used to model the ion distribution functions in plasma and fusion experiments { \cite{Salewski2018,Moseev2019}, for fast ions generated through wave heating in the ion cyclotron range of frequencies or neutral beam injection.}

In this work, we derive two- and one-dimensional integral forms of $\langle\sigma v\rangle$, which combine the existing results for Maxwellian (thermal), beam (drift), and bi-Maxwellian (anisotropic) distributions, using drift bi-Maxwellian distributions. This will provide valuable input for the modeling of fast ion fusion yields in numerical codes, such as NUBEAM \cite{Pankin2004}.

In Section \ref{sec:eq}, we present the derivations. In Section \ref{sec:apply}, we explore the potential enhancement of reactivity due to beam and temperature anisotropies and examine the modification to the Lawson fusion gain criteria. Finally, in Section \ref{sec:summ}, we summarize our results.

\section{Fusion Reactivity for Drift bi-Maxwellian Ion Velocity Distributions}\label{sec:eq}

Our goal is to derive the one-dimensional integral form of $\langle\sigma v\rangle$ for the most general reactant distribution functions $f_1$ and $f_2$. { In this section, only the main results will be summarized, with all the details of the derivations and proofs provided in the supplemental document.}

\subsection{Several existing results}\label{sec:sgmvexist}

In this subsection, we summarize several existing results for reference.

When considering two Maxwellian ions/reactants with distribution functions given by
\begin{equation}
f_j({\bm v})=\Big(\frac{m_j}{2\pi k_BT_j}\Big)^{3/2}\exp\Big(-\frac{m_jv^2}{2k_BT_j}\Big),
\end{equation}
where $j=1, 2$, and $k_B$ is the Boltzmann constant, it is well-known \cite{Nevins2000} that the integral form of $\langle\sigma v\rangle$ is
\begin{equation}\label{eq:sgmvm}
\langle\sigma v\rangle_M=\sqrt{\frac{8}{\pi m_r}}\frac{1}{(k_BT_{r})^{3/2}}\int_0^{\infty}\sigma(E)E\exp\Big(-\frac{E}{k_BT_{r}}\Big)dE,
\end{equation}
where $T_{r}$ is an effective temperature defined as
\begin{equation}
T_{r}=\frac{m_1T_2+m_2T_1}{m_1+m_2}.
\end{equation}
It is important to note that the temperatures of the two reactants $T_1$ and $T_2$ are not necessarily equal, and increasing the temperature of the lighter reactant has a more significant impact on reactivity compared to increasing the temperature of the heavier reactant{, which is due to the larger increase of the relative velocity between two reactants.}

For two drift Maxwellian ions, the velocity distribution function is given by
\begin{equation}\label{eq:fbm}
f_j({\bm v})=\Big(\frac{m_j}{2\pi k_BT_j}\Big)^{3/2}\exp\Big[-\frac{m_j({\bm v}-{\bm v}_{dj})^2}{2k_BT_j}\Big].
\end{equation}
After some derivations (similar to those in Sec.\ref{subsec:dbm}), we find that
\begin{eqnarray}\label{eq:sgmvdm}\nonumber
&&\langle\sigma v\rangle_{DM}\\\nonumber
&=&\frac{2}{\sqrt{\pi}v_{tr}v_{d}}\int_0^{\infty}\sigma(v)v^2\exp\Big(-\frac{v^2+v_{d}^2}{v_{tr}^2}\Big)\cdot\\\nonumber
&&\sinh\Big(2\frac{vv_{d}}{v_{tr}^2}\Big)dv\\\nonumber
&=&\sqrt{\frac{2}{\pi m_rk_B^2T_rT_d}}\int_0^{\infty}\sigma(E)\sqrt{E}\cdot\\
&&\exp\Big(-\frac{E+E_{d}}{k_BT_r}\Big)\sinh\Big(\frac{2\sqrt{EE_d}}{k_BT_r}\Big)dE,
\end{eqnarray}
where $\sinh(x) = (e^x - e^{-x})/2 \simeq x + x^3/6 + \cdots$, the effective temperature $T_{r}$, thermal velocity $v_{tr}$, drift velocity $v_{d}$, and drift energy $E_{d}$ are defined as follows
\begin{eqnarray}\nonumber
T_{r}=\frac{m_1T_2+m_2T_1}{m_1+m_2},~~v_{tr}=\sqrt{\frac{2k_BT_r}{m_r}},\\
v_{d}=|{\bm v}_{d2}-{\bm v}_{d1}|, ~~E_d\equiv k_BT_d=\frac{m_rv_{d}^2}{2}.
\end{eqnarray}
Note that the drift velocity ${\bm v}_{dj}$ can be in arbitrary directions, i.e., ${\bm v}_{d1}$ and ${\bm v}_{d2}$ are not required to be in the same direction. The result described by Eq.(\ref{eq:sgmvdm}) is more general than those in other literature, such as Refs.\cite{Miley1974,Miley1975,Morse2018,Ou2015,Mikkelsen1989,Li2022}. For example, for $T_{1} \to 0$ and $v_{d2} = 0$, the result reduces to a beam-target case\cite{Miley1974,Miley1975,Morse2018}. If considering only the beam in the same direction, Eq.(\ref{eq:sgmvdm}) can be reduced to the results in \cite{Appelbe2011,Ou2015,Li2022}. If $E_{d} = 0$, Eq.(\ref{eq:sgmvdm}) reduces to Eq.(\ref{eq:sgmvm}).

For two bi-Maxwellian ions, the velocity distribution function is given by
\begin{eqnarray}\label{eq:fdbm}
f_j({\bm v})=\frac{1}{T_{\parallel j}^{1/2}T_{\perp j}}\Big(\frac{m_j}{2\pi k_B}\Big)^{3/2}\exp\Big(-\frac{m_jv_{\perp}^2}{2k_BT_{\perp j}}-\frac{m_jv_{\parallel}^2}{2k_BT_{\parallel j}}\Big).
\end{eqnarray}
It has been shown in Refs. \cite{Kolmes2021, Li2022} that the reactivity can be reduced to
\begin{eqnarray}\nonumber\label{eq:sgmvbm}
\langle\sigma v\rangle_{BM}&=&\Big(\frac{2}{\pi m_rk_B^3}\Big)^{1/2}\frac{1}{{T_{\parallel r}^{1/2}T_{\perp r}}}\int_0^{\infty}dE_\perp\int_0^{\infty}dE_{\parallel}\cdot\\\nonumber
&&\sigma(E)\sqrt{\frac{E}{E_\parallel}}\exp\Big[-\frac{E_\parallel}{{k_B}T_{\parallel r}}-\frac{E_\perp}{{k_B}T_{\perp r}}\Big]\\\nonumber
&=&\sqrt{\frac{2}{m_rk_B^2T_{\perp r}(T_{\perp r}-T_{\parallel r})}}\int_0^{\infty}dE\sigma(E){\sqrt{E}}\cdot\\
&&\exp\Big(-\frac{E}{k_BT_{\perp r}}\Big){\rm erf}\Big[\sqrt{\frac{E(T_{\perp r}-T_{\parallel r})}{k_B T_{\perp r}T_{\parallel r}}}\Big],
\end{eqnarray}
where ${\rm erf}(x)=\frac{2}{\sqrt{\pi}}\int_0^{x}e^{-t^2}dt\simeq\frac{2}{\sqrt{\pi}}(x-\frac{x^3}{3}+\cdots)$ is the error function, and the effective parallel and perpendicular temperatures and energies are defined as
\begin{eqnarray}\nonumber
T_{\parallel r}=\frac{m_1T_{\parallel 2}+m_2T_{\parallel 1}}{m_1+m_2},~~T_{\perp r}=\frac{m_1T_{\perp 2}+m_2T_{\perp 1}}{m_1+m_2},\\
E=E_\parallel+E_\perp,~~E_\parallel=\frac{1}{2}m_rv_\parallel^2,~~E_\perp=\frac{1}{2}m_rv_\perp^2.
\end{eqnarray}
To calculate $\langle\sigma v\rangle_{BM}$ for $T_{\perp r} < T_{\parallel r}$, we can use the relation ${\rm erf}(ix) = i \cdot {\rm erfi}(x)$ to keep the argument of the error function a real number, with ${\rm erfi}(x) = \frac{2}{\sqrt{\pi}}\int_0^{x}e^{t^2}dt$. For $T_{\perp r} = T_{\parallel r}$, Eq.(\ref{eq:sgmvbm}) reduces to Eq.(\ref{eq:sgmvm}). It should be noted that as $x \to +\infty$, ${\rm erf}(x) \simeq 1 - \frac{e^{-x^2}}{\sqrt{\pi}}(\frac{1}{x} - \frac{1}{2x^3} + \frac{3}{4x^5} + \cdots)$ and ${\rm erf}(-x) = -{\rm erf}(x)$.

In equations (\ref{eq:sgmvm}), (\ref{eq:sgmvdm}), and (\ref{eq:sgmvbm}), the subscripts 'M', 'DM', and 'BM' respectively denote Maxwellian, drift Maxwellian, and bi-Maxwellian distributions.

\begin{figure*}
%\centering
\includegraphics[width=18cm]{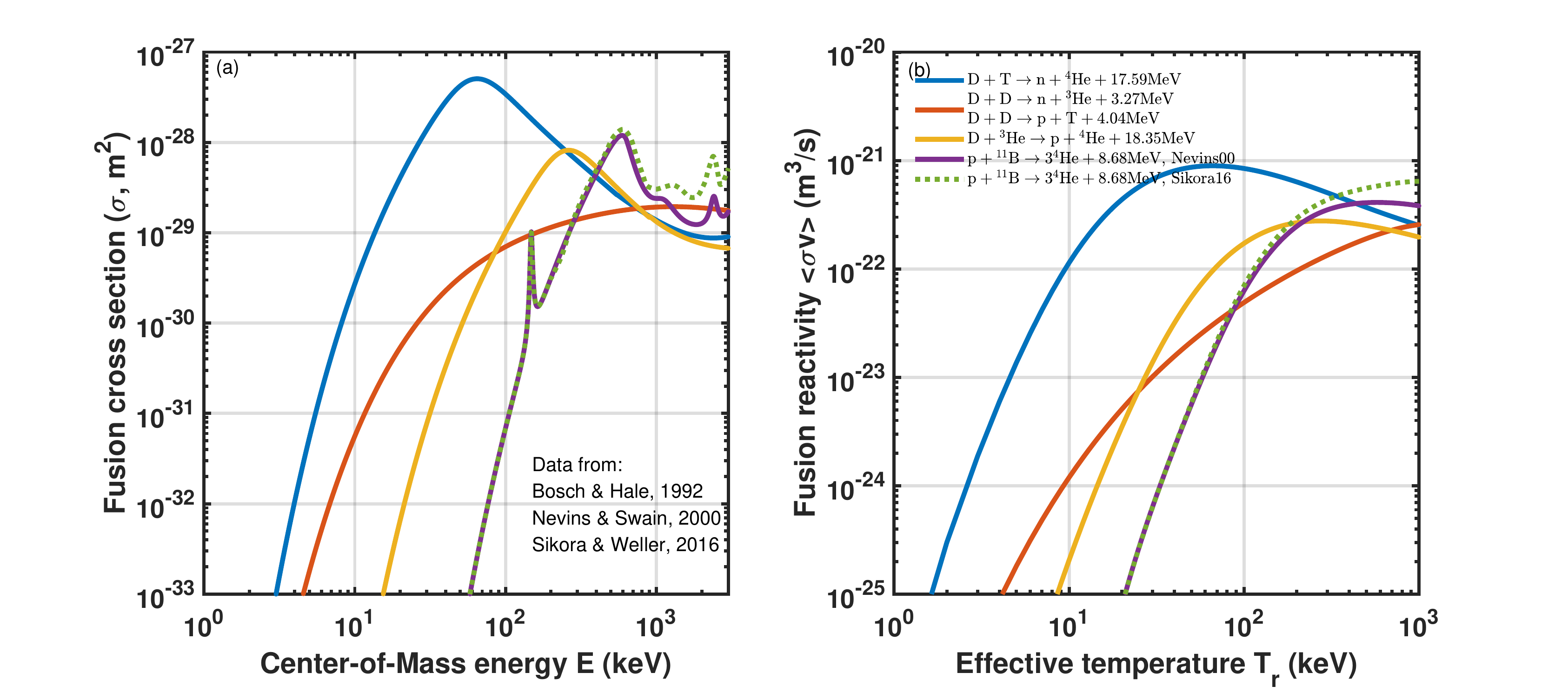}
\caption{Fusion cross-sections and Maxwellian fusion reactivities for several of the most important fusion energy-relevant reactions, including D-T, D-D, D-He3, and p-B11.}\label{fig:fusion_cross_section}
\end{figure*}

\subsection{Drift bi-Maxwellian ions}\label{subsec:dbm}

In principle, if we were to use drift tri-Maxwellian distribution functions, the results for $\langle\sigma v\rangle$ would contain all of the results from Subsection \ref{sec:sgmvexist}. However, only a three-dimensional integral form has been obtained \cite{Nath2013}, and it is not straightforward to obtain a one-dimensional form.

As a result, in this analysis, we will limit ourselves to two drift bi-Maxwellian ions. The distribution functions are given by
\begin{eqnarray}\nonumber
f_j({\bm v_j})&=&\frac{1}{T_{\parallel j}^{1/2}T_{\perp j}}\Big(\frac{m_j}{2\pi k_B}\Big)^{3/2}\cdot\\
&&\exp\Big[-\frac{m_jv_{\perp j}^2}{2k_BT_{\perp j}}-\frac{m_j(v_{\parallel j}-v_{dj})^2}{2k_BT_{\parallel j}}\Big],
\end{eqnarray}
with $\int f_j({\bm v_j}) d{\bm v}_j=1$, $v_{\perp j}^2=v_{xj}^2+v_{yj}^2$, and $v_{\parallel j}=v_{zj}$. Here, unlike in Eq.(\ref{eq:fbm}), we limit the drift velocities ${\bm v}_{dj}=v_{djz}{\hat{\bm z}}$ to be in only the parallel direction, which is typically parallel to the magnetic field ${\bm B}=B{\hat{\bm z}}$. 

Using the following transformation
\begin{eqnarray}\nonumber
{\bm v}={\bm v}_1-{\bm v}_2,~~m_r\equiv\frac{m_1m_2}{m_1+m_2},~~E=\frac{1}{2}m_rv^2,\\\nonumber
T_{\parallel r}=\frac{m_1T_{\parallel 2}+m_2T_{\parallel 1}}{m_1+m_2},~~T_{\perp r}=\frac{m_1T_{\perp 2}+m_2T_{\perp 1}}{m_1+m_2},\\\nonumber
{\bm v}_c=\frac{m_2T_{\perp 1}{\bm v}_{\perp 2}+m_1T_{\perp 2}{\bm v}_{\perp 1}}{{m_1T_{\perp 2}}+m_2T_{\perp 1}}+\\\frac{m_2T_{\parallel1}({\bm v}_{\parallel 2}-{\bm v}_{d2})+m_1T_{\parallel 2}({\bm v}_{\parallel 1}-{\bm v}_{d1})}{{m_1T_{\parallel 2}}+m_2T_{\parallel 1}},
\end{eqnarray}
we have the Jacobian $J=|d{\bm v}_1d{\bm v}_2/d{\bm v}_cd{\bm v}|=1$, and 
\begin{eqnarray}\nonumber
&&\frac{m_1(v_{\parallel 1}-v_{d1})^2}{T_{\parallel 1}}+\frac{m_2(v_{\parallel 2}-v_{d2})^2}{T_{\parallel 2}}\\\nonumber&&=\Big(\frac{m_1}{T_{\parallel 1}}+\frac{m_2}{T_{\parallel 2}}\Big){\bm v}_{c\parallel}^2+\frac{m_r}{T_{\parallel r}}v_{d}^2+\frac{m_r}{T_{\parallel r}}{\bm v}_\parallel^2+2\frac{m_r}{T_{\parallel r}}{\bm v}_\parallel v_{d}.
\end{eqnarray}
Thus, we have
\begin{eqnarray}\label{eq:sgmvdbm3d}\nonumber
&&\langle\sigma v\rangle_{DBM}\\\nonumber
&=&\int\int d{\bm v}_1d{\bm v}_2\sigma(|{\bm v}_1-{\bm v}_2|)|{\bm v}_1-{\bm v}_2|f_1({\bm v}_1)f_2({\bm v}_2)\\\nonumber
&=&\Big(\frac{m_r}{2\pi k_B}\Big)^{3/2}\frac{1}{T_{\parallel r}^{1/2}T_{\perp r}}\int d{\bm v}\sigma(v)v\cdot\\
&&\exp\Big[-\frac{m_rv_{\perp}^2}{2k_BT_{\perp r}}-\frac{m_r(v_{\parallel}+v_{d})^2}{2k_BT_{\parallel r}}\Big].
\end{eqnarray}
Therefore, the six-dimensional integral is reduced to a three-dimensional one. A similar treatment can be found in reference \cite{Nath2013}.

We then use the following definitions
\begin{eqnarray}\nonumber
E=E_\parallel+E_\perp,~~E_\parallel=\frac{1}{2}m_rv_\parallel^2,~~E_\perp=\frac{1}{2}m_rv_\perp^2,\\\nonumber
v_{d}=v_{d2}-v_{d1}, ~~E_d=k_BT_d=\frac{m_rv_{d}^2}{2},\\\nonumber
v_x=\sqrt{2E_\perp/m_r}\cos\phi,~~v_y=\sqrt{2E_\perp/m_r}\sin\phi,\\\nonumber
v_z=\pm\sqrt{2E_\parallel/m_r},
\end{eqnarray}
where {$\phi\in[0,2\pi]$}, $E_\perp\in[0,\infty)$ and $E_\parallel\in[0,\infty)$. This results in
\begin{eqnarray}\nonumber
d{\bm v}=\frac{1}{m_r^{3/2}}\sqrt{\frac{1}{2E_\parallel}}dE_\parallel dE_\perp d\phi.
\end{eqnarray}
Thus, equation (\ref{eq:sgmvdbm3d}) can be rewritten as a two-dimensional integral
\begin{eqnarray}\label{eq:sgmvdbm2d}\nonumber
\langle\sigma v\rangle_{DBM}&=&{\Big(\frac{1}{2\pi m_rk_B^3}\Big)^{1/2}}\frac{1}{T_{\parallel r}^{1/2}T_{\perp r}}\int_{0}^{\infty}\int_{0}^{\infty} dE_\parallel dE_\perp\sigma(E)\cdot\\\nonumber
&&\sqrt{\frac{E}{E_\parallel}}\exp\Big[-\frac{E_{\perp}}{k_BT_{\perp r}}-\frac{(E_{\parallel}+E_d)}{k_BT_{\parallel r}}\Big]\cdot\\
&&\Big[\exp\Big(-2\frac{\sqrt{E_{\parallel} E_d}}{k_BT_{\parallel r}}\Big)+\exp\Big(2\frac{\sqrt{E_{\parallel} E_d}}{k_BT_{\parallel r}}\Big)\Big].
\end{eqnarray}

We can proceed further with another transformation, given by
\begin{eqnarray}\nonumber
E=E_\parallel+E_\perp,~~t^2=\frac{E_\parallel(T_{\perp r}-T_{\parallel r})}{k_BT_{\perp r}T_{\parallel r}},
\end{eqnarray}
or,
\begin{eqnarray}\nonumber
E_\parallel=t^2\frac{k_BT_{\perp r}T_{\parallel r}}{{(T_{\perp r}-T_{\parallel r})}},~~E_\perp=E-t^2\frac{k_BT_{\perp r}T_{\parallel r}}{{(T_{\perp r}-T_{\parallel r})}},\\\nonumber
dE_\parallel dE_\perp=2t\frac{k_BT_{\perp r}T_{\parallel r}}{{(T_{\perp r}-T_{\parallel r})}}dEdt.
\end{eqnarray}
With this, we can finally arrive at a one-dimensional integral form of fusion reactivity for drift bi-Maxwellian reactants, as follows
\begin{eqnarray}\label{eq:sgmvdbm1d}\nonumber
&&\langle\sigma v\rangle_{DBM}=\\\nonumber
&&\exp\Big[\frac{E_d}{k_B(T_{\perp r}-T_{\parallel r})}\Big]\sqrt{\frac{1}{2m_rk_B^2T_{\perp r}(T_{\perp r}-T_{\parallel r})}}\cdot\\\nonumber
&&\int_0^{\infty}dE\sigma(E)\sqrt{E}\exp\Big(-\frac{E}{k_BT_{\perp r}}\Big)\cdot\\\nonumber
&&\Big[{\rm erf}\Big(\sqrt{E\frac{(T_{\perp r}-T_{\parallel r})}{k_BT_{\parallel r}T_{\perp r}}}+\sqrt{\frac{T_{\perp r}E_d}{k_BT_{\parallel r}(T_{\perp r}-T_{\parallel r})}}\Big)+\\
&&{\rm erf}\Big(\sqrt{E\frac{(T_{\perp r}-T_{\parallel r})}{k_BT_{\parallel r}T_{\perp r}}}-\sqrt{\frac{T_{\perp r}E_d}{k_BT_{\parallel r}(T_{\perp r}-T_{\parallel r})}}\Big)\Big].
\end{eqnarray}
The above result holds for both $T_{\perp r}>T_{\parallel r}$ and $T_{\perp r}<T_{\parallel r}$, if the imaginary error function ${\rm erf}(ix)=i\cdot {\rm erfi}(x)$ is used. The subscript `DBM' in the result stands for drift bi-Maxwellian.

Both equations (\ref{eq:sgmvdbm1d}) and (\ref{eq:sgmvdbm2d}) can be used for practical numerical calculations, as the computation costs for the 1D and 2D integrals are not excessive. Equation (\ref{eq:sgmvdbm1d}) can be easily reduced to equation (\ref{eq:sgmvbm}) when $E_d=0$, and can be reduced to equation (\ref{eq:sgmvdm}) when $T_{\parallel r}=T_{\perp r}$ { (as detailed in the supplemental document)}. Fast approximate formulations can be used to calculate the error functions ${\rm erf}(x)$ and ${\rm erfi}(x)$. Care should be taken when the two reactants are the same, for example, in D-D (Deuterium-Deuterium) fusion, the drift $E_d$ should be treated as two groups of reactants (representing the two groups of D ions colliding with each other) to avoid double-counting. In practical situations, the velocity distribution of one species can be constructed using multiple drift bi-Maxwellian distributions, and the calculation of the fusion reactivity would still be straightforward using the above formulations.

\begin{figure}
\centering
\includegraphics[width=9cm]{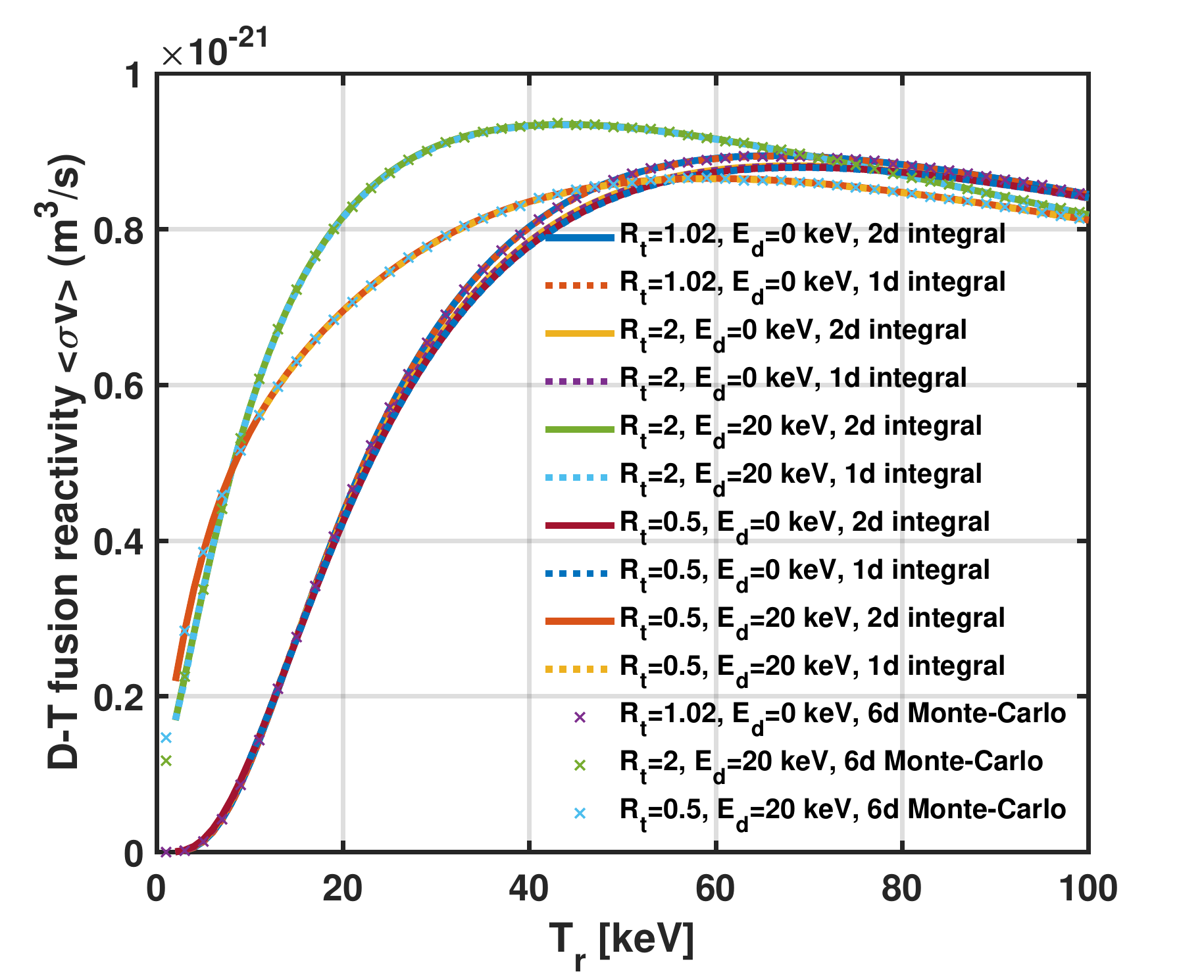}
\caption{D-T fusion reactivities with drift bi-Maxwellian distribution ions. The results of the 1D integral Eq.(\ref{eq:sgmvdbm1d}) are identical to those of the 2D integral Eq.(\ref{eq:sgmvdbm2d}), {and also agree with the 6D Monte-Carlo integral based on Eq.(\ref{eq:sgmv}).}}\label{fig:sgmvdbmdt_1d_2d_mc_RtEd}
\end{figure}

\begin{figure}
\centering
\includegraphics[width=9cm]{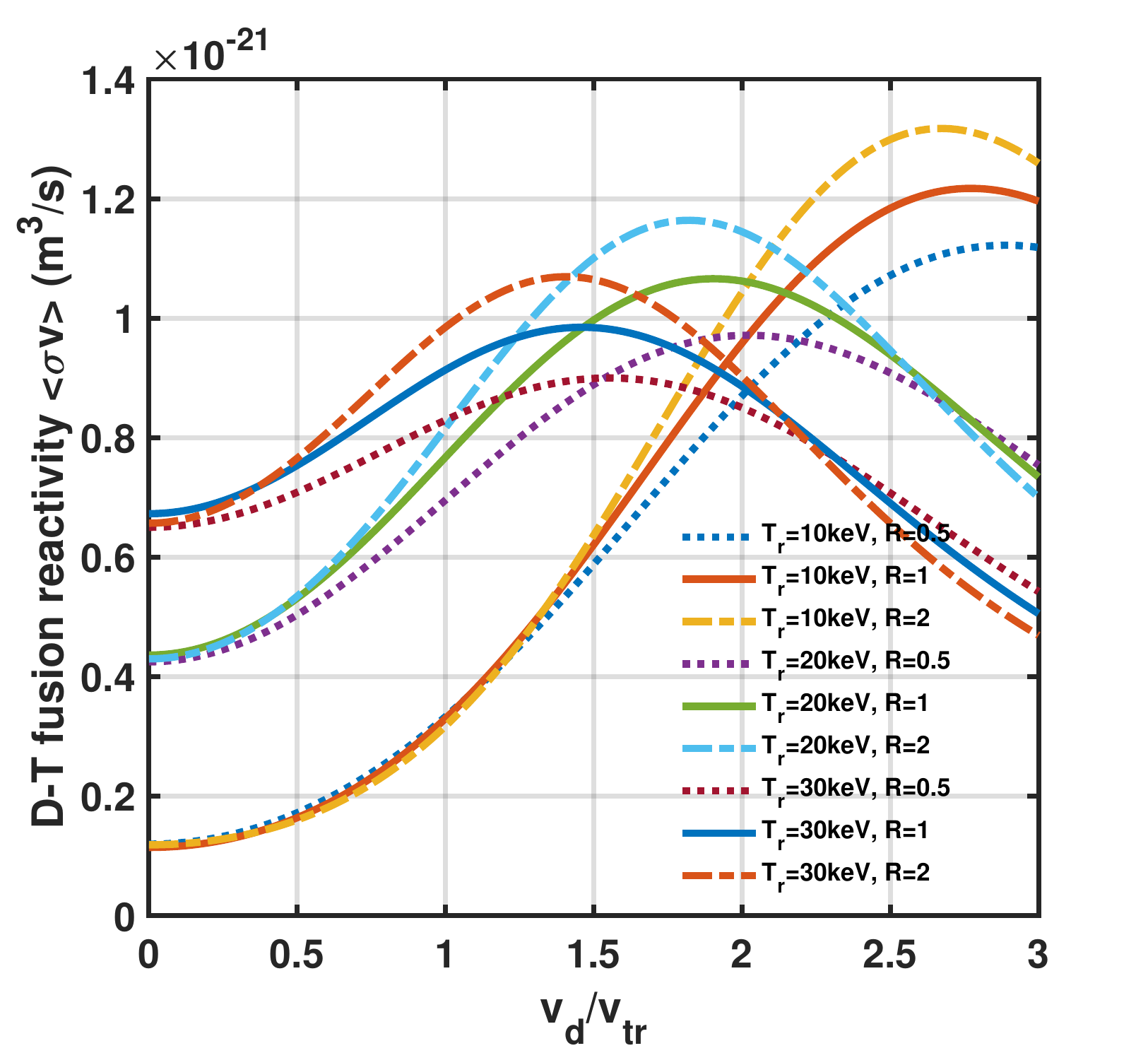}
\caption{D-T fusion reactivities with drift bi-Maxwellian distribution ions are presented, considering different temperatures ($T_r=10,~20,~30$ keV), anisotropics ($R=0.5,~1.0,~2.0$), and drift velocities ($v_d/v_{tr} \in [0,3]$). The results are found to be in close agreement with Ref. \cite{Nath2013} Figure 4.}\label{fig:bench_nath13}
\end{figure}

\subsection{Analytical insights}

The one-dimensional integral form of the fusion reactivity, $\langle\sigma v\rangle$, provides analytical insight into the effect of the reactants velocity distribution functions, $f_{1,2}$, on the fusion reactivity. To this end, we can define a kernel distribution function, $K(E)$, as follows
\begin{eqnarray}\label{eq:sgmvk}
\langle\sigma v\rangle=\frac{1}{\sqrt{m_r}}\int_0^{\infty}\sigma(E)K(E)dE.
\end{eqnarray}
$K(E)$ can be easily obtained from Eqs. (\ref{eq:sgmvm}), (\ref{eq:sgmvdm}), (\ref{eq:sgmvbm}) and (\ref{eq:sgmvdbm1d}) for Maxwellian, drift Maxwellian, bi-Maxwellian, and drift bi-Maxwellian distributions, respectively. For example, for a Maxwellian distribution
\begin{eqnarray}\label{eq:sgmvk}
K_M(E)=\sqrt{\frac{8}{\pi }}\frac{1}{(k_BT_{r})^{3/2}}E\exp\Big(-\frac{E}{k_BT_{r}}\Big).
\end{eqnarray}

The fusion reactivity, $\langle\sigma v\rangle$, is determined by the overlap between the kernel function, $K(E)$, and the peak regions of the cross section, $\sigma(E)$. In this sense, $K(E)$ can be viewed as a weighted function of the cross section, $\sigma(E)$. Typical kernel functions, $K(E)$, will be displayed in Sec. \ref{sec:apply} (Fig. \ref{fig:KE}).

\begin{figure*}
\centering
\includegraphics[width=18cm]{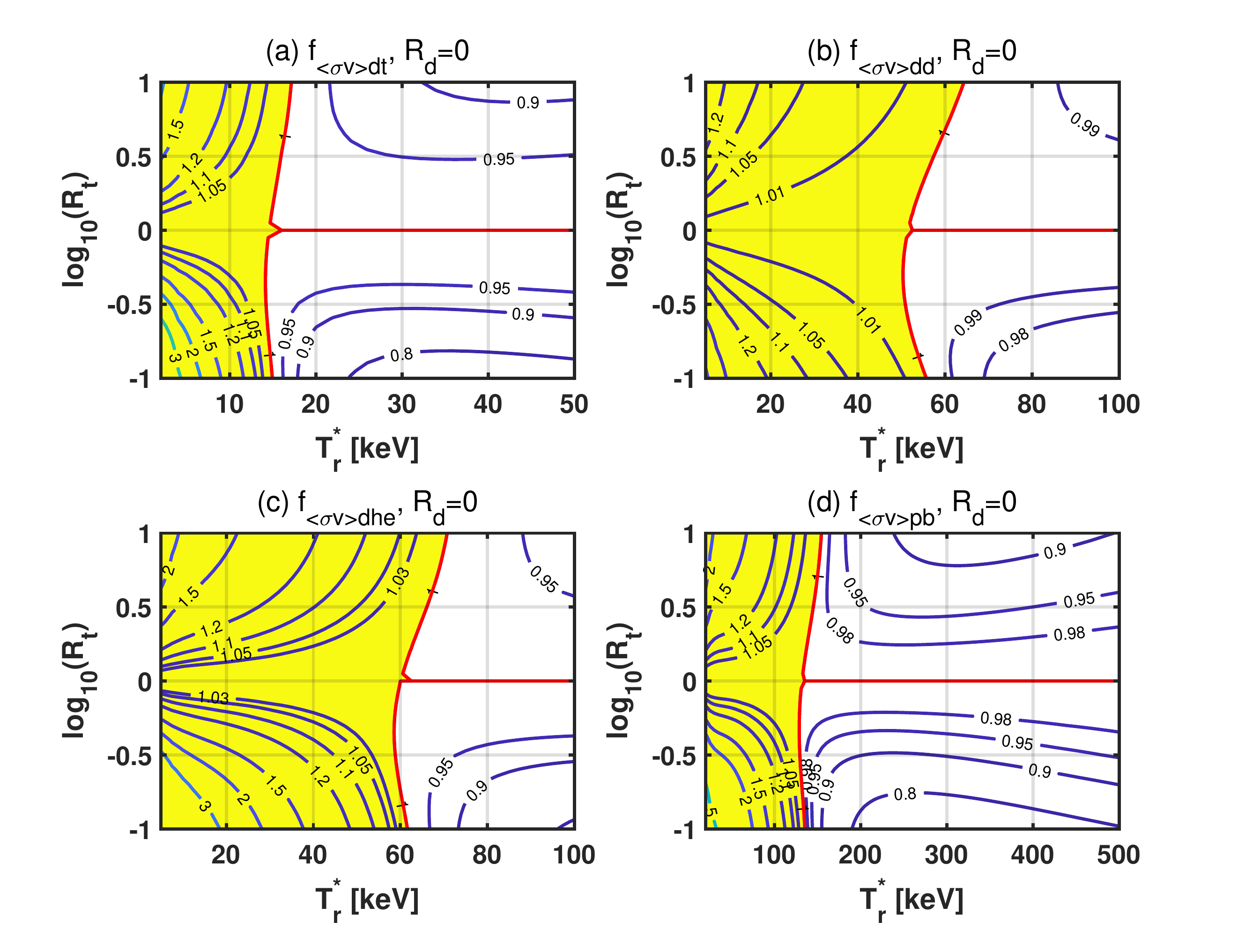}
\caption{The enhancement factor $f_{\langle\sigma v\rangle}$ for fusion reactivities of { D-T, D-D, D-He3, and p-B11 reactions} is shown as a function of temperature anisotropics ($R=T_{\perp r}/T_{\parallel r}$) for different temperatures. The shading in the figure indicates regions where $f_{\langle\sigma v\rangle}>1$, implying that the fusion reactivity is enhanced.}\label{fig:f_Tr_R}
\end{figure*}

\begin{figure}
\centering
\includegraphics[width=9cm]{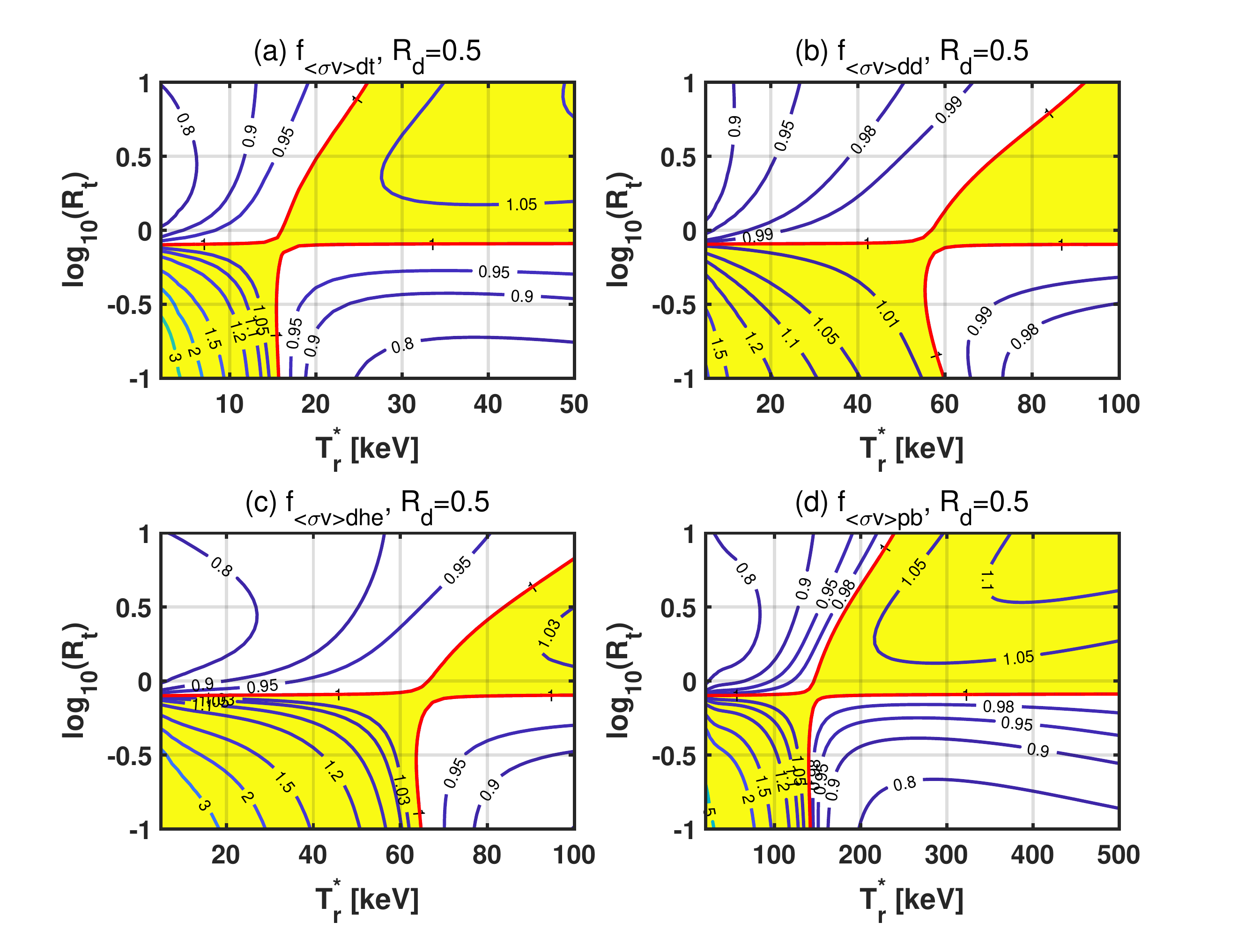}
\caption{Fusion reactivities enhancement factor $f_{\langle\sigma v\rangle}$ { for D-T, D-D, D-He3, and p-B11 reactions}, with temperature anisotropics ($R=T_{\perp r}/T_{\parallel r}$) and parallel drift for different effective kinetic energies $k_BT_r^*=\frac{2}{3}E_{kr}$, with $R_d=0.5$. { The shading range indicates where $f_{\langle\sigma v\rangle}>1$.}}\label{fig:f_Tr_Rt_Rd=0.5}
\end{figure}

\section{Applications}\label{sec:apply}

\subsection{Benchmarks}

For fusion energy studies, the most notable fusion reactions are ${\rm D-T}$ (Deuterium-Trillium), ${\rm D-D}$, ${\rm D-{}^3He}$ (Helium) and ${\rm p-^{11}B}$ (proton-Boron), due to their relatively large fusion cross sections. The fusion cross sections and corresponding reactivities with their reactants in Maxwellian distributions are shown in Fig.\ref{fig:fusion_cross_section}. The cross section data for ${\rm D-T}$, ${\rm D-D}$, ${\rm D-{}^3He}$ were taken from Ref.\cite{Bosch1992}, while the data for ${\rm p-^{11}B}$ were taken from Refs.\cite{Nevins2000} and \cite{Sikora2016}. Note that the $\sigma(E)$ data are typically only available for $E<4$ MeV. Care should be taken when using $\langle\sigma v\rangle$ results at high energy ranges, and the convergence of the calculations should be checked.

\begin{figure}
\centering
\includegraphics[width=9cm]{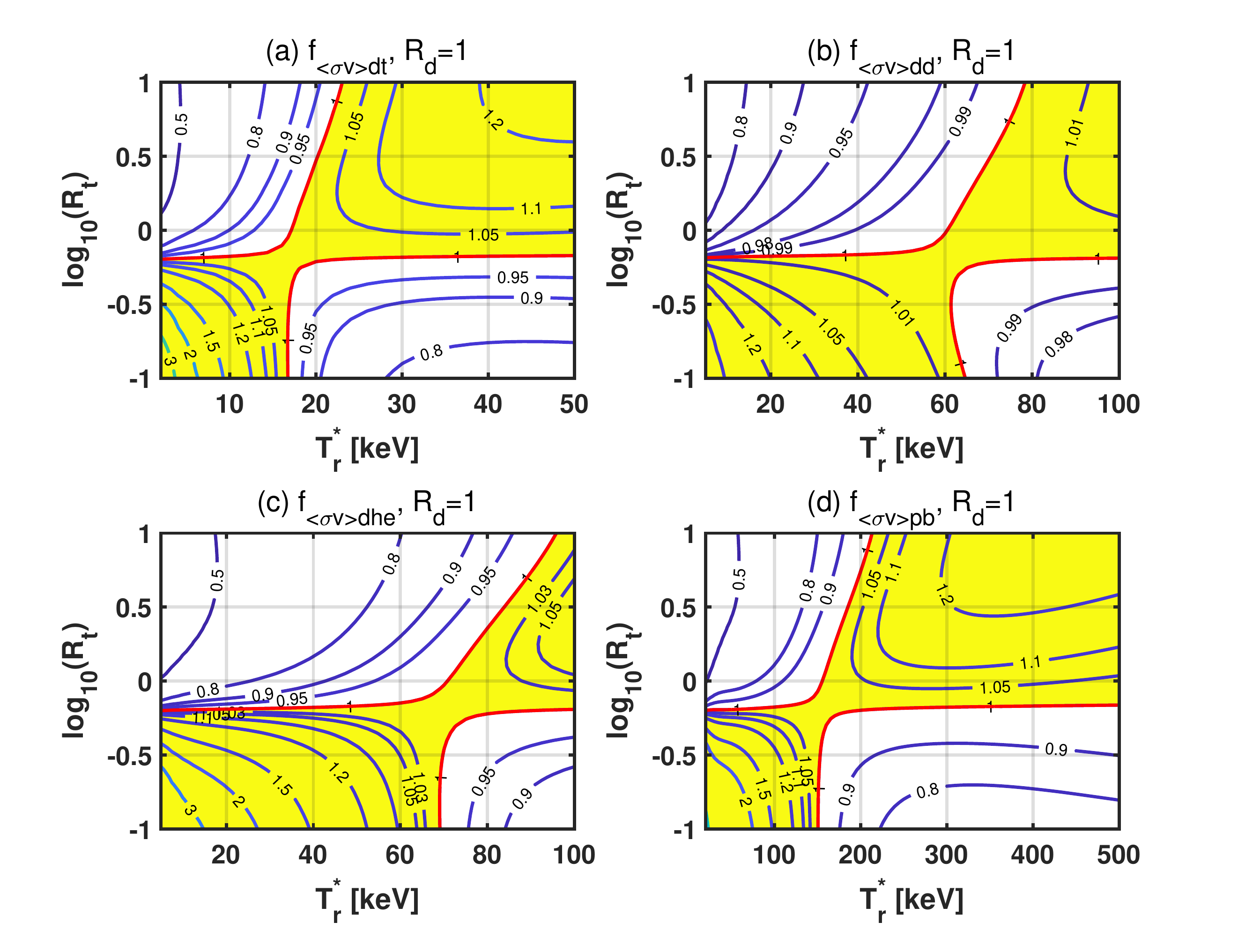}
\caption{Fusion reactivities enhancement factor $f_{\langle\sigma v\rangle}$ { for D-T, D-D, D-He3, and p-B11 reactions}, with temperature anisotropics ($R=T_{\perp r}/T_{\parallel r}$) and parallel drift for different effective kinetic energies $k_BT_r^*=\frac{2}{3}E_{kr}$, with $R_d=1$. { The shading range indicates where $f_{\langle\sigma v\rangle}>1$.}}\label{fig:f_Tr_Rt_Rd=1}
\end{figure}

To validate Eq. (\ref{eq:sgmvdbm1d}), we compare its numerical integrals with those of Eqs. (\ref{eq:sgmvdbm2d}) { and (\ref{eq:sgmv})}. Figure \ref{fig:sgmvdbmdt_1d_2d_mc_RtEd} demonstrates that the results of the 1D integral of Eq. (\ref{eq:sgmvdbm1d}) are consistent with those of the 2D integral of Eq. (\ref{eq:sgmvdbm2d}) and { the {6D Monte Carlo \cite{Xie2023}} integral of Eq. (\ref{eq:sgmv})} for all three cases: $R_t=T_{\perp r}/T_{\parallel r}\simeq 1$, $>1$ and $<1$.

We also compare our results to those in Ref. \cite{Nath2013}. Figure \ref{fig:bench_nath13} shows the D-T fusion reactivities for drift bi-Maxwellian distribution ions with different temperatures ($T_r=10,~20,~30$ keV), anisotropies ($R=R_t=0.5,~1.0, ~2.0$) and drift velocities ($v_d/v_{tr}\in[0,3]$). The results are in good agreement with the 3D integral in Fig. 4 of Ref. \cite{Nath2013}. However, it should be noted that our normalization parameter, $v_{tr}=\sqrt{2/m_r}/\sqrt{3/m_D}v_{rms}\simeq1.054v_{rms}$, is different from the normalization parameter in Ref. \cite{Nath2013}. The latter only studied drift velocities of $v_{d}/v_{rms}\leq1$ and concluded that the drift always enhances the reactivities. However, our results show that for large drift velocities, such as $v_{d}/v_{rms}\simeq3$, the enhancement factor can decrease and become less than 1 for $T_r=30$ keV. The reason for this can be found by comparing the trends in the variation of the fusion cross section in Fig. \ref{fig:fusion_cross_section} with the kernel function $K_{DBM}(E)$. For large beam drift, the center of mass energy $E$ is larger than the peak cross section energy, which decreases the fusion reactivity.

\begin{figure}
\centering
\includegraphics[width=9cm]{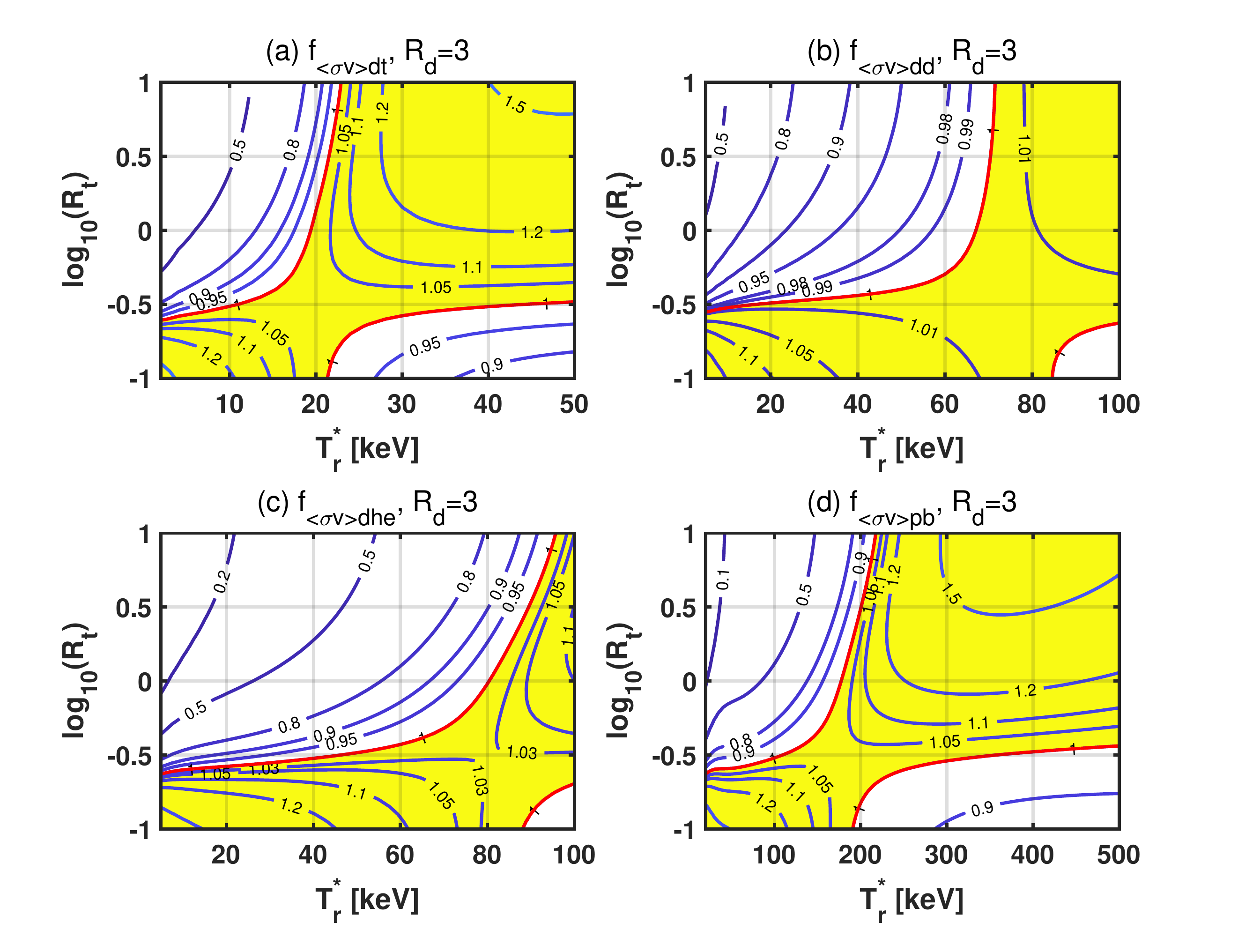}
\caption{Fusion reactivities enhancement factor $f_{\langle\sigma v\rangle}$ { for D-T, D-D, D-He3, and p-B11 reactions}, with temperature anisotropics ($R=T_{\perp r}/T_{\parallel r}$) and parallel drift for different effective kinetic energies $k_BT_r^*=\frac{2}{3}E_{kr}$, with $R_d=3$. The shading range indicates where $f_{\langle\sigma v\rangle}>1$.}\label{fig:f_Tr_Rt_Rd=3}
\end{figure}

\begin{figure}
\centering
\includegraphics[width=9cm]{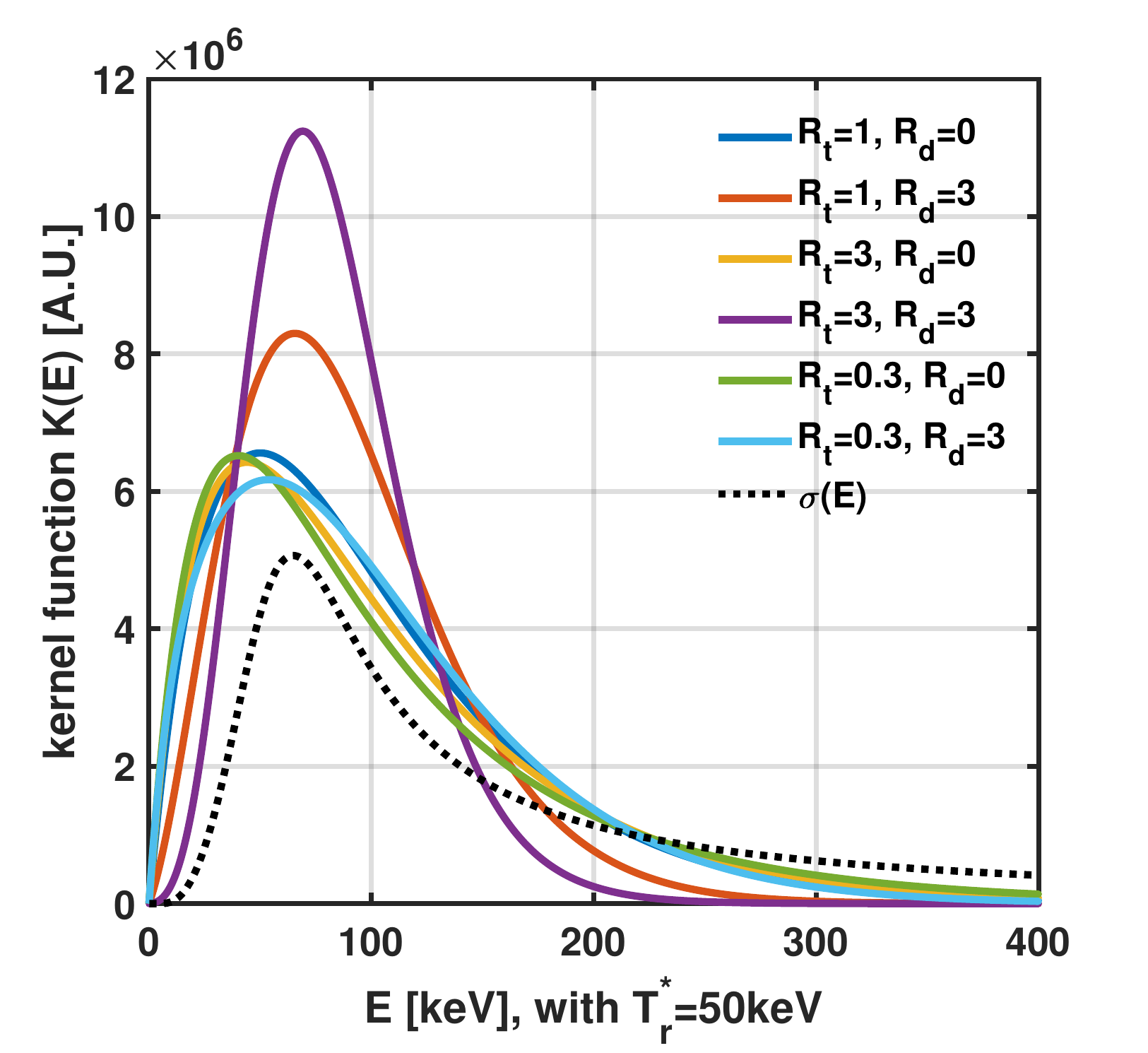}
\caption{Kernel functions $K(E)$ and D-T fusion cross sections $\sigma(E)$ are shown for typical different drift bi-Maxwellian distributions with anisotropies $R_t$ and drift $R_d$.}\label{fig:KE}
\end{figure}

\subsection{{Potential for Increased Reactivity}}

Although the physics of a possible reactivity enhancement due to the synergy of beam effect and temperature anisotropies can be studied via the 3D integral formulation in Ref. \cite{Nath2013}, the 2D formulation given by Eq. (\ref{eq:sgmvdbm2d}) and the 1D formulation given by Eq. (\ref{eq:sgmvdbm1d}) are much simpler and more efficient for calculations, and can also be used to evaluate the impact of each term.

To investigate the possibility of a reactivity enhancement, we fix some parameters. For $E_d=0$, we keep the total kinetic energy constant, resulting in the same $T_r$. However, for $E_d\neq0$, the situation is different. The average kinetic energy is calculated from the distribution function as $E_k=\frac{1}{2}m\int v^2 f({\bm v})d{\bm v}$. For the drift bi-Maxwellian as described in Eq.(\ref{eq:fdbm}), the average kinetic energy is given by
\begin{eqnarray}\label{eq:Ekj}
E_{kj}=\frac{k_BT_{\parallel j}+2k_BT_{\perp j}}{2}+\frac{1}{2}m_jv_{dj}^2,
\end{eqnarray}
where $E_{kj}$ is the sum of the thermal energy, $E_{th,j}=\frac{k_BT_{\parallel j}+2k_BT_{\perp j}}{2}$, and the drift energy, $E_{dj}=\frac{1}{2}m_jv_{dj}^2$. Equation (\ref{eq:Ekj}) was derived by evaluating the velocity space integrals using the formulas $\int_{-\infty}^{\infty}v^2\exp[-(v-v_d)^2/a^2]dv=\sqrt{\pi}a(2v_d^2 + a^2)/2$ and $\int_{-\infty}^{\infty}\exp[-(v-v_d)^2/a^2]dv=\sqrt{\pi}a$.

It is assumed that the masses satisfy $m_1\leq m_2$, meaning the first reactant is lighter than the second. The fusion reactivity $\langle\sigma v\rangle_{DBM}$ depends only on $m_r$, $T_{\perp r}$, $T_{\parallel r}$, and $E_d$. To minimize the total kinetic energy $E_k$, it is optimal to choose the first reactant as the one that is both drift and thermal, i.e., $v_{d2}\ll v_{d1}$ and $T_{2}\ll T_1$.

To study the reactivity enhancement, it is important to keep the total kinetic energy, $n_1E_{k1} + n_2E_{k2}$, constant. In this study, to simplify the discussion, we keep the average effective kinetic energy constant, defined as $E_{kr} = (k_BT_{\parallel r} + 2k_BT_{\perp r})/2 + E_d$. To facilitate the analysis, we introduce new variables 
\begin{eqnarray}
T_r=\frac{(2T_{\perp r}+T_{\parallel r})}{3}, ~R_t=\frac{T_{\perp r}}{T_{\parallel r}},~R_d=\frac{E_d}{k_BT_{r}}.
\end{eqnarray}
The thermal energy is kept constant for a fixed $T_r$, and the average effective kinetic energy is kept constant for a fixed $E_{kr}$. Hence, we study the relation between the fusion reactivity $\langle\sigma v\rangle_{DBM}$ and $E_{kr}$, $R_t$, and $R_d$. We also define the fusion reactivity enhancement factor as 
\begin{eqnarray}
f_{\langle\sigma v\rangle}\equiv\frac{{\langle\sigma v\rangle}}{{\langle\sigma v\rangle}_{R_t=1,R_d=0}}, 
\end{eqnarray} 
where $E_{kr}$ is kept constant. Finally, we introduce a new temperature $T_r^{*}\equiv 2E_{kr}/(3k_B)$, which reduces to $T_r^{*} = T_r$ for $E_d = 0$.

Figure \ref{fig:f_Tr_R} shows the fusion reactivity enhancement factors for different temperatures with only temperature anisotropy considered. It can be seen that for all four of these fusion reactions, the enhancements only occur at low temperatures, such as the critical temperatures for D-T, D-D, D-He3, and p-B11, which are around 15 keV, 50 keV, 60 keV, and 140 keV, respectively. Similar conclusions have been drawn in previous studies, as reported in Refs. \cite{Kalra1988, Nath2013, Kolmes2021, Li2022}. The reason for this can be found by examining the cross section plot (Fig. \ref{fig:fusion_cross_section}) and the kernel function $K(E)$. The enhancements occur when the effects of the cross section peak are maximized. For the D-D fusion reactivity, the anisotropic effects are weak and contribute less than 5\% in most ranges. The effects are slightly larger for D-He3 compared to D-D. For D-T and p-B11, changes greater than 5\% can easily be achieved. Since the enhancements are small in most cases, the results reported in Ref. \cite{Mirin1983} are understandable, which concluded that the fusion reactivity in TFTR affected by the distortion of bulk ions from Maxwellians is small.

Figures \ref{fig:f_Tr_Rt_Rd=0.5}, \ref{fig:f_Tr_Rt_Rd=1}, and \ref{fig:f_Tr_Rt_Rd=3} show the fusion reactivity enhancement factors with both temperature anisotropy and parallel drift, for $R_d=0.5,~1,~3$, respectively. It can be observed that the effects on the D-D and D-He3 fusion reactions are small. The p-B11 result is calculated using the cross-section data from Ref. \cite{Nevins2000}. In comparison to the case without drift, the case with drift can also increase the reactivity for $T_{\perp r}\gtrsim T_{\parallel r}$ with a large $T_r^{*}$. Among all cases, in the range of $R_t\lesssim1$ and low $T_r^{*}$, the fusion reactivity can be enhanced, even by over 100\%. These situations commonly occur in modern tokamak experiments, such as TFTR, by using neutral beam injection (NBI) or the ion cyclotron range of frequencies (ICRF) \cite{Hawryluk1998}.

We are particularly interested in the p-B11 fusion, which is typically considered difficult as a fusion energy source \cite{Nevins2000}. In Fig. \ref{fig:f_Tr_Rt_Rd=3}, at $T_r^*\simeq250$ keV and $T_{\perp r}>0.75T_{\parallel r}$, the enhancement can reach around $20\%$. Although this enhancement factor is not large, its effects on the fusion energy gain factor could be significant, as demonstrated in subsection \ref{sec:lawson}. Figure \ref{fig:KE} shows the kernel functions $K(E)$ and the D-T fusion cross-section $\sigma(E)$ for typical different values of $R_t$ and $R_d$. { It is readily apparent that the overlap of the kernel function $K(E)$ with the peak of the cross-section $\sigma(E)$ is largest for the case $(R_t=3, R_d=3)$, indicating that the fusion reactivity for this case should be the largest among the six cases in the figure. By comparing the results in Figs. \ref{fig:f_Tr_R} and \ref{fig:f_Tr_Rt_Rd=3}, we see that this is indeed the case.}

\subsection{Impact on the Lawson criteria}\label{sec:lawson}

The fusion reactivity enhancement factors are variable for different parameters. In order to show the impact on the Lawson fusion condition\cite{Lawson1957,Wurzel2022}, we choose a fixed $f_{\langle\sigma v\rangle}$. The fusion gain factor $Q$ is defined as $Q\equiv(P_{out}-P_{in})/P_{in}$, where $P_{in}$ is the input power and $P_{out}$ is the output power. A scientific breakeven occurs when $Q=1$ and ignition occurs when $Q=\infty$. For a steady fusion reactor, we can calculate $Q$ using the following equation
\begin{eqnarray}\label{eq:Q}
Q=\frac{P_{fus}}{E_{th}/\tau_E-f_{ion}P_{fus}+P_{rad}},
\end{eqnarray}
where $\tau_E$ is the energy confinement time, and the thermal energy and fusion power are given by
\begin{eqnarray}
E_{th}&=&\frac{3}{2}k_B\sum_jn_jT_j=\frac{3}{2}k_B(n_eT_e+n_iT_i),\\
P_{fus}&=&\frac{1}{1+\delta_{12}}n_1n_2\langle\sigma v\rangle Y.
\end{eqnarray}
Here, $Y$ is the energy release of each fusion reaction, and $Y_{+}$ is the energy to charged products, with $f_{ion}=Y_{+}/Y$. The densities of the two ions, $n_1$ and $n_2$, are given as $n_1=x_1n_i$ and $n_2=x_2n_i$, respectively, where $x_1$ and $x_2$ are the ion concentrations and $n_i=n_1+n_2$. If the two ions are not the same, then $x_1=x$ and $x_2=1-x$. The quasi-neutrality condition requires the electron density, $n_e$, to be equal to $n_e=Z_1n_1+Z_2n_2=Z_in_i$, where $Z_1$ and $Z_2$ are the charge numbers of the two ions. The effective charge, $Z_{eff}$, is given by $Z_{eff}=[x_1Z_1^2+x_2Z_2^2]/[x_1Z_1+x_2Z_2]$.

We consider only the weak relativistic bremsstrahlung radiation losses $P_{rad}=P_{brem}$, with
\begin{eqnarray}
P_{brem}&=&C_Bn_e^2\sqrt{k_BT_e}Z_{eff}g_{eff}~{\rm [MW\cdot m^{-3}]},\\\nonumber
g_{eff}&=&\Big[1+0.7936\frac{k_BT_e}{m_ec^2}+1.874\Big(\frac{k_BT_e}{m_ec^2}\Big)^2\Big]\\
&&+\frac{1}{Z_{eff}}\frac{3}{\sqrt{2}}\frac{k_BT_e}{m_ec^2},
\end{eqnarray}
in which $C_B=5.34\times10^{-37}$. The unit for $k_BT_e$ is keV, for density $n_e$ is ${\rm m}^{-3}$.

We can finally obtain the Lawson criteria
\begin{eqnarray}
n_e\tau_E=\frac{\frac{3}{2}k_B(T_e+T_i/Z_i)}{\frac{(1/Q+f_{ion})}{1+\delta_{12}}\frac{x_1x_2}{Z_i^2}\langle\sigma v\rangle Y -C_B\sqrt{k_BT_e}Z_{eff}g_{eff}}.
\end{eqnarray}

\begin{figure*}
\centering
\includegraphics[width=18cm]{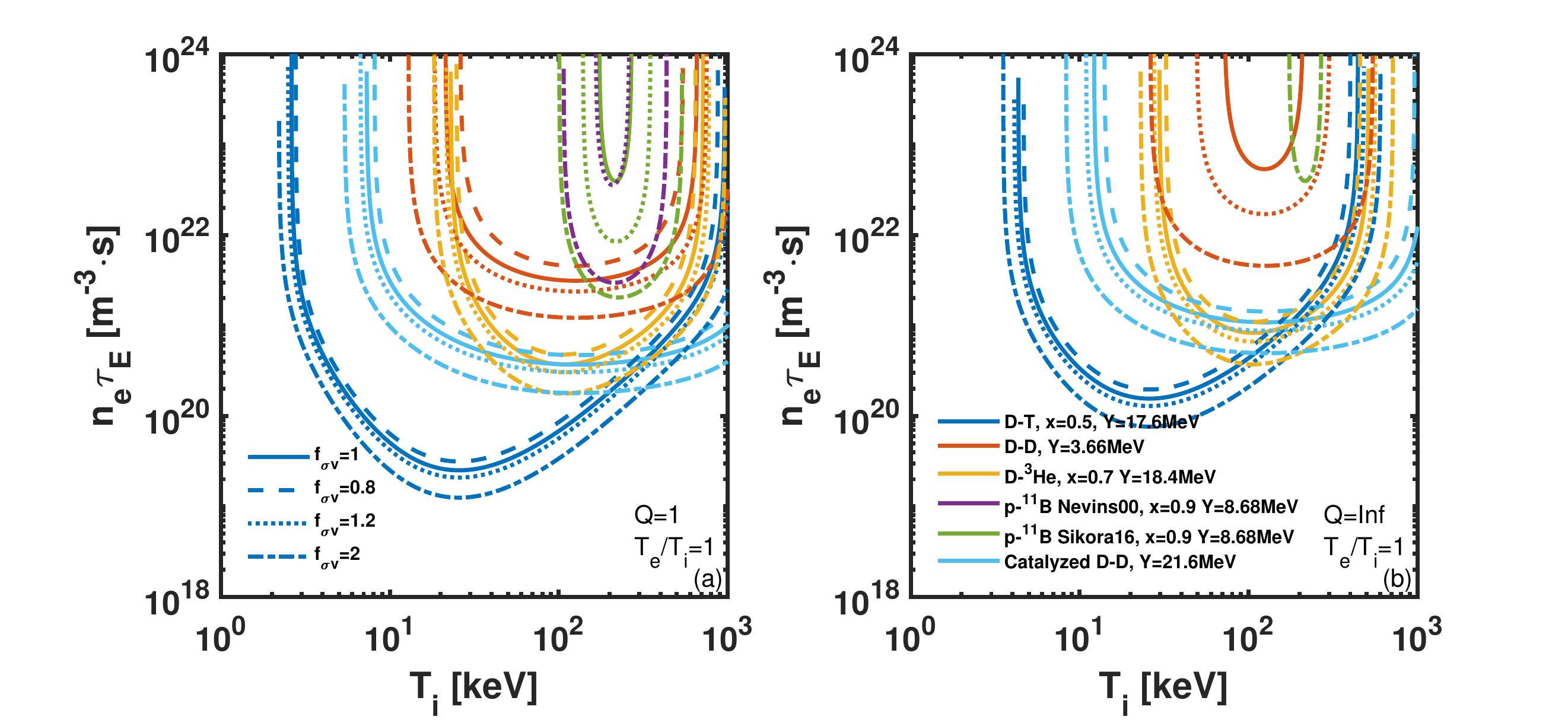}
\caption{The effect of varying the fusion reactivity enhancement factor, $f_{\langle\sigma v\rangle}$, with values of 0.8, 1.0, 1.2, and 2.0, on the Lawson criteria of D-T, D-D, D-He3, p-B11, and catalyzed D-D fusions is shown for the cases of $Q=1$ (left) and $Q=\infty$ (right), with the assumption of equal electron and ion temperatures, $T_e=T_i$. Note that some p-B11 lines may be invisible in the figure.}\label{fig:Lawson_ntauET}
\end{figure*}

\begin{figure*}
\centering
\includegraphics[width=18cm]{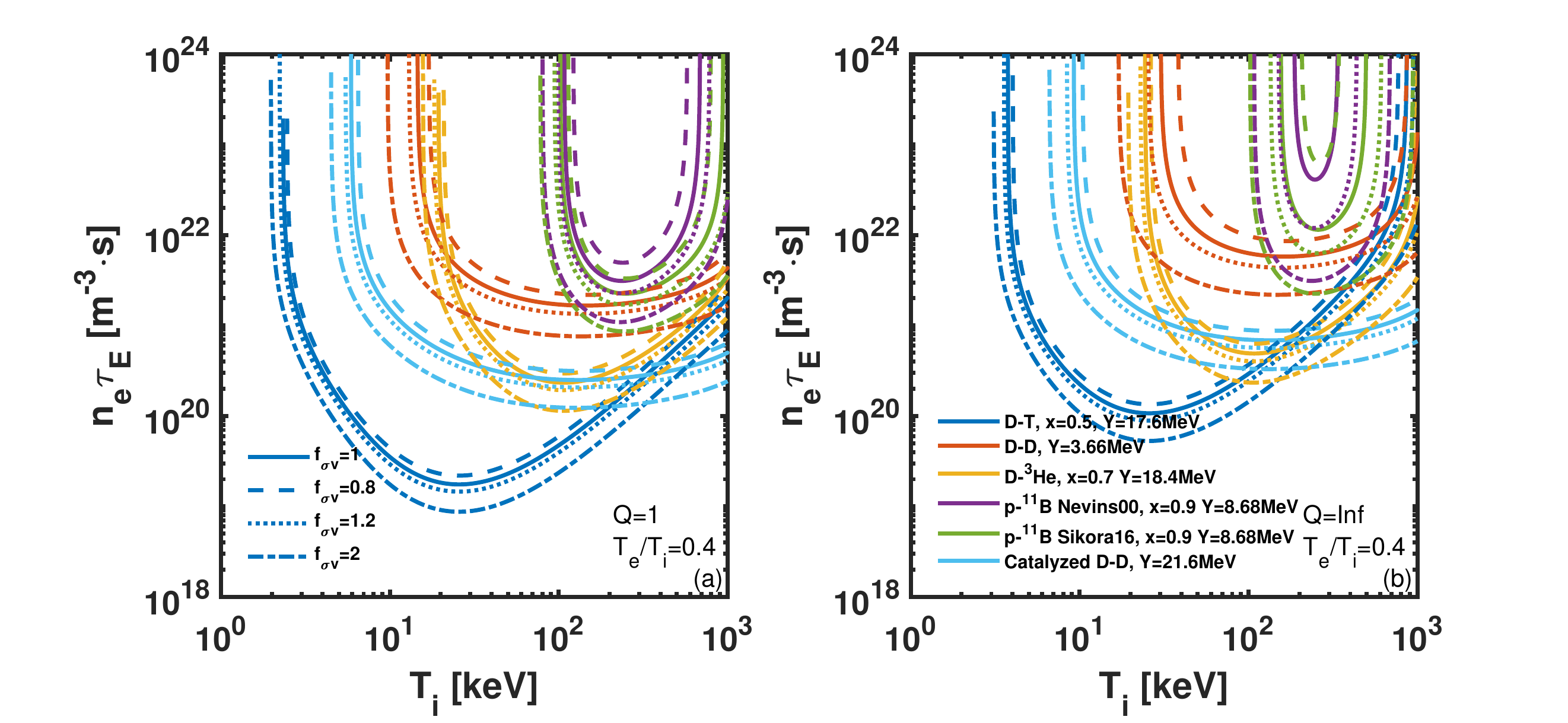}
\caption{The effect of varying the fusion reactivity enhancement factor, $f_{\langle\sigma v\rangle}$, with values of 0.8, 1.0, 1.2, and 2.0, on the Lawson criteria of D-T, D-D, D-He3, p-B11, and catalyzed D-D fusions is shown for the cases of $Q=1$ (left) and $Q=\infty$ (right), with the assumption of hot ion mode, $T_e/T_i=0.4$. Note that some p-B11 lines may be invisible in the figure.}\label{fig:Lawson_ntauET_fT=0.4}
\end{figure*}

Here, we consider $\langle\sigma v\rangle = f_{\langle\sigma v\rangle} \cdot \langle\sigma v\rangle_M$, and we would like to see how much the enhancement factor $f_{\langle\sigma v\rangle}$ can affect the Lawson criteria. Figure \ref{fig:Lawson_ntauET} illustrates the effect of changing the fusion reactivity enhancement factor $f_{\langle\sigma v\rangle} = 0.8,~1.0,~1.2,~2.0$ on the Lawson criteria of D-T, D-D, D-He3, p-B11, and catalyzed D-D fusions, for both $Q=1$ and $Q=\infty$. The enhancement factor $f_{\langle\sigma v\rangle}$ affects the Lawson criteria for all these fusion reactions, with the most significant effect seen for the p-B11 fusion. For the catalyzed D-D fusion, we use the same fusion reactivity as D-D fusion, but with a larger yield energy of $Y=21.6$ MeV instead of $Y=3.66$ MeV. In Fig. \ref{fig:Lawson_ntauET}, considering the p-B11 cross section by Nevins et al. \cite{Nevins2000} with $f_{\langle\sigma v\rangle} = 1$, or the cross section by Sikora et al. \cite{Sikora2016} with $f_{\langle\sigma v\rangle} = 0.8$, self-sustaining burning is not supported by the Lawson criteria ($Q\geq1$ is impossible). With a small change in $f_{\langle\sigma v\rangle}$ from 1.0 to 1.2, the Lawson condition derived from the cross section of Sikora et al. \cite{Sikora2016} shows a decline of an order of magnitude. This is also the main reason why the self-sustaining burning of p-B11 was previously thought to be impossible \cite{Nevins1998}, but later reversed \cite{Putvinskia2019} with new data that only enhanced the fusion cross section in the high energy regime ($>0.5$ MeV). This sensitivity is mainly due to the close balance between the power of the radiation loss and the fusion energy release, with $P_{fus}/P_{rad}\simeq1$.

If we cannot increase the p-B11 fusion reactivity, as indicated by the cross section \cite{Sikora2016} and distribution functions in this work, a hot-ion mode with $T_i/T_e > 1$ may be required for p-B11 fusion to achieve energy gain \cite{Cai2022}. As shown in Fig. \ref{fig:Lawson_ntauET_fT=0.4}, even with the fusion reactivity enhancement factor, a hot-ion mode with $T_e/T_i=0.4$ and { all other parameters being the same as in Fig. \ref{fig:Lawson_ntauET}}, does not significantly relax the Lawson criteria for p-B11 fusion. Thus, reducing the radiation loss is more feasible than enhancing the fusion reactivity.

\section{Summary and Conclusion}\label{sec:summ}

The formulation of the fusion reactivity for two drift bi-Maxwellian reactants has been derived into both two- and one-dimensional integral forms. This formulation combines previous results in the literature, including Maxwellian-Maxwellian, beam-target, and bi-Maxwellian reactants. The results show that enhancement factors in fusion reactivity due to temperature anisotropies and parallel drift can be greater than 20\% at temperatures relevant to fusion energy. This enhancement is particularly important for the p-B11 fusion reaction. To maximize reactivity enhancement, the distribution function should remain close to the peak region of the cross section, which can be understood from the kernel function $K(E)$ in the one-dimensional fusion reactivity integral formulation. This means that the reactivity can only be enhanced within specific ranges. For all four fusion reactions (D-T, D-D, D-He3, and p-B11), the ranges that benefit the most from reactivity enhancement without considering drift are $T < T_c$, where $T_c$ represents critical temperatures of $T_c^{DT} \simeq 15$ keV, $T_c^{DD} \simeq 50$ keV, $T_c^{DHe3} \simeq 60$ keV, and $T_c^{pB11} \simeq 140$ keV. When drift is taken into account, enhancement also occurs when $T > T_c$ if $T_{\perp r} \gtrsim T_{\parallel r}$. Results show that even a modest increase in fusion reactivity can lower the Lawson criteria for p-B11 fusion dramatically. Thus, the impact of the distribution function on the fusion reactivity is crucial in the study of p-B11 fusion. Previous studies\cite{Rostoker2005,Peng2020} of p-B11 fusion with beam fusion should be carefully examined, as they treated drift energy oversimplified. A more practical study would keep the total energy ($n_1E_{k1}+n_2E_{k2}$) constant, which could provide more insight into the best choice for $T_{\parallel,\perp j}$ and $v_{dj}$, and lead to further results based on those shown in section \ref{sec:apply}. It should be noted that the beam and anisotropy can also affect the confinement time $\tau_E$, which has not been discussed here.

The present work could also be useful in improving the ion energy spectrum diagnostic method, as discussed in Refs.\cite{Appelbe2011,Li2022a}. Furthermore, it is interesting to consider if there are other integral forms of the fusion reactivity that can be obtained in two- or one-dimensional, such as suprathermal distributions\cite{Majumdar2011,Majumdar2016}, fast ion tails\cite{Niikura1990,Goncharov2015}, and drift bi-Maxwellian distributions with perpendicular drift and ring drift\cite{Xie2019}. These could be topics for future research.

\ack HSX would like to express his gratitude for the valuable discussions with Yang Li at the early stage of this project.
{We would also like to thank the valuable comments and suggestions from the two anonymous referees.}



\section*{Supplementary material}

# The Derivations of the Fusion Reactivity for Drift bi-Maxwellian Distributions

Huasheng XIE, huashengxie@gmail.com, ENN

Update: 2023-02-05

This is supplemental document for the paper "Huasheng Xie, Muzhi Tan, Di Luo, Zhi Li and Bing Liu, Fusion Reactivities with Drift bi-Maxwellian Ion Velocity Distributions, 2022", which gives the details of the derivations of the fusion reactivity for Maxwellian, drift Maxwellian, bi-Maxwellian and drift bi-Maxwellian distributions.

Fusion reactivity

$$\langle \sigma v \rangle = \iint d\boldsymbol{v}_1 d\boldsymbol{v}_2 \sigma(v) v f_1(\boldsymbol{v}_1) f_2(\boldsymbol{v}_2),$$

where

$$v = |\boldsymbol{v}| = |\boldsymbol{v}_1 - \boldsymbol{v}_2|.$$

and

$$v_x = v\sin\theta\cos\phi, v_y = v\sin\theta\sin\phi, v_z = v\cos\theta.$$
$$\theta = [0,\pi], \phi = [0,2\pi], v = [0, \infty), d\boldsymbol{v} = v^2 \sin\theta\, dv\, d\theta\, d\phi.$$

## 1. Maxwellian distributions

For Maxwellian distributions

$$f_j(\boldsymbol{v}_j) = \left(\frac{m_j}{2\pi k_B T_j}\right)^{\frac{3}{2}} \exp\left(-\frac{m_j v_j^2}{2k_B T_j}\right),$$

$$\int d\boldsymbol{v}\, f_j(\boldsymbol{v}_j) = 1,$$

we define

$$\boldsymbol{v} = \boldsymbol{v}_1 - \boldsymbol{v}_2, \quad \boldsymbol{v}_c = \frac{m_1 T_2 \boldsymbol{v}_1 + m_2 T_1 \boldsymbol{v}_2}{m_1 T_2 + m_2 T_1}, \quad m_r \equiv \frac{m_1 m_2}{m_1 + m_2}, \quad E = \frac{1}{2} m_r v^2,$$

$$T_r \equiv \frac{m_2 T_1 + m_1 T_2}{m_1 + m_2}, \quad \frac{m_1 m_2}{T_1 T_2} = \left(\frac{m_1}{T_1} + \frac{m_2}{T_2}\right)\frac{m_r}{T_r},$$

i.e.,

$$\boldsymbol{v}_1 = \boldsymbol{v}_c + \frac{m_2 T_1}{m_1 T_2 + m_2 T_1} \boldsymbol{v},$$

$$\boldsymbol{v}_2 = \boldsymbol{v}_c - \frac{m_1 T_2}{m_1 T_2 + m_2 T_1} \boldsymbol{v},$$

gives

$$d\boldsymbol{v}_1 d\boldsymbol{v}_2 = d\boldsymbol{v} d\boldsymbol{v}_c,$$

where we have used the below Jacobian

$$J = \frac{d\boldsymbol{v}_1 d\boldsymbol{v}_2}{d\boldsymbol{v} d\boldsymbol{v}_c} = \frac{\partial(v_{1x}, v_{1y}, v_{1z}, v_{2x}, v_{2y}, v_{2z})}{\partial(v_x, v_y, v_z, v_{cx}, v_{cy}, v_{cz})} = \begin{vmatrix} a & & & 1 & & \\ & a & & & 1 & \\ & & a & & & 1 \\ b & & & 1 & & \\ & b & & & 1 & \\ & & b & & & 1 \end{vmatrix} = 1,$$

$$a = \frac{m_2 T_1}{m_1 T_2 + m_2 T_1}, b = -\frac{m_1 T_2}{m_1 T_2 + m_2 T_1}.$$

We can have

$$\langle \sigma v \rangle_M = \iint d\boldsymbol{v}_1 d\boldsymbol{v}_2 \sigma(v) v f_1(\boldsymbol{v}_1) f_2(\boldsymbol{v}_2) =$$

$$= \left(\frac{m_1 m_2}{(2\pi k_B)^2 T_1 T_2}\right)^{\frac{3}{2}} \iint d\boldsymbol{v}_1 d\boldsymbol{v}_2 \sigma(v) v \exp\left(-\frac{m_1 v_1^2}{2 k_B T_1} - \frac{m_2 v_2^2}{2 k_B T_2}\right)$$

$$= \left(\frac{m_1 m_2}{(2\pi k_B)^2 T_1 T_2}\right)^{\frac{3}{2}} \iint d\boldsymbol{v} d\boldsymbol{v}_c \sigma(v) v \exp\left(-\frac{m_1 v_1^2}{2 k_B T_1} - \frac{m_2 v_2^2}{2 k_B T_2}\right).$$

And since

$$\frac{m_1 v_1^2}{T_1} + \frac{m_2 v_2^2}{T_2} = \frac{m_1}{T_1}\left[(v_{cx}+av_x)^2 + (v_{cy}+av_y)^2 + (v_{cz}+av_z)^2\right]$$

$$+ \frac{m_2}{T_2}\left[(v_{cx}+bv_x)^2 + (v_{cy}+bv_y)^2 + (v_{cz}+bv_z)^2\right]$$

$$= \frac{m_1}{T_1}\left[v_c^2 + a^2 v^2 + 2av_{cx}v_x + 2av_{cy}v_y + 2av_{cz}v_z\right]$$

$$+ \frac{m_2}{T_2}\left[v_c^2 + b^2 v^2 + 2bv_{cx}v_x + 2bv_{cy}v_y + 2bv_{cz}v_z\right]$$

$$= \left(\frac{m_1}{T_1} + \frac{m_2}{T_2}\right) v_c^2$$

$$+ \left[\frac{m_1}{T_1}\left(\frac{m_2 T_1}{m_1 T_2 + m_2 T_1}\right)^2 + \frac{m_2}{T_2}\left(\frac{m_1 T_2}{m_1 T_2 + m_2 T_1}\right)^2\right] v^2$$

$$+ 2\left(\frac{m_1}{T_1}\frac{m_2 T_1}{m_1 T_2 + m_2 T_1} - \frac{m_2}{T_2}\frac{m_1 T_2}{m_1 T_2 + m_2 T_1}\right)\left[v_{cx}v_x + v_{cy}v_y + v_{cz}v_z\right]$$

$$= \left(\frac{m_1}{T_1} + \frac{m_2}{T_2}\right) v_c^2 + \frac{m_1 m_2}{m_1 T_2 + m_2 T_1}\ v^2 = \left(\frac{m_1}{T_1} + \frac{m_2}{T_2}\right) v_c^2 + \frac{m_r}{T_r}\ v^2.$$

We have

$$\langle\sigma v\rangle_M = \left(\frac{m_1 m_2}{(2\pi k_B)^2 T_1 T_2}\right)^{\frac{3}{2}} \iint d\boldsymbol{v}d\boldsymbol{v}_c \sigma(v) v \exp\left(-\frac{m_1 v_1^2}{2k_B T_1} - \frac{m_2 v_2^2}{2k_B T_2}\right)$$

$$= \left(\frac{1}{2\pi k_B}\left(\frac{m_1}{T_1}\right.\right.$$

$$\left.\left.+ \frac{m_2}{T_2}\right)\right)^{\frac{3}{2}} \left(\frac{m_r}{2\pi k_B T_r}\right)^{\frac{3}{2}} \iint d\boldsymbol{v}d\boldsymbol{v}_c \sigma(v) v \exp\left(-\frac{1}{2k_B}\left(\frac{m_1}{T_1}+\frac{m_2}{T_2}\right)v_c^2\right.$$

$$\left.- \frac{1}{2k_B}\frac{m_r}{T_r}\ v^2\right) == \left(\frac{m_r}{2\pi k_B T_r}\right)^{\frac{3}{2}} \int d\boldsymbol{v}\sigma(v)v \exp\left(-\frac{1}{2k_B}\frac{m_r}{T_r}\ v^2\right)$$

$$= \left(\frac{m_r}{2\pi k_B T_r}\right)^{\frac{3}{2}} \int_0^\infty 4\pi v^2 dv \exp\left(-\frac{m_r}{2k_B T_r}v^2\right)\sigma(v)v$$

$$= \left(\frac{8}{\pi m_r}\right)^{\frac{1}{2}}\left(\frac{1}{k_B T_r}\right)^{\frac{3}{2}} \int_0^\infty \exp\left(-\frac{E}{k_B T_r}\right)\sigma(E) E dE,$$

The above result is the same as that in Nevins00.

## 2. Drift Maxwellian distributions

For drift Maxwellian distribution

$$f_j(v_j) = \left(\frac{m_j}{2\pi k_B T_j}\right)^{\frac{3}{2}} \exp\left(-\frac{m_j v_{\perp j}^2}{2k_B T_j}\right) \exp\left(-\frac{m_j(v_{\|j} - v_{dzj})^2}{2k_B T_j}\right),$$

$$v_{\perp j}^2 = v_{xj}^2 + v_{yj}^2, \qquad v_{\|j} = v_{zj},$$

we define

$$\boldsymbol{v} = \boldsymbol{v}_1 - \boldsymbol{v}_2, \qquad \boldsymbol{v}_c = \frac{m_1 T_2(\boldsymbol{v}_1 - \boldsymbol{v}_{dz1}) + m_2 T_1(\boldsymbol{v}_2 - \boldsymbol{v}_{dz2})}{m_1 T_2 + m_2 T_1},$$

$$m_r \equiv \frac{m_1 m_2}{m_1 + m_2}, \qquad E \equiv \frac{1}{2} m_r v^2,$$

$$T_r \equiv \frac{m_2 T_1 + m_1 T_2}{m_1 + m_2}, \qquad \frac{m_1 m_2}{T_1 T_2} = \left(\frac{m_1}{T_1} + \frac{m_2}{T_2}\right)\frac{m_r}{T_r},$$

$$v_{tr} \equiv \sqrt{\frac{2k_B T_r}{m_r}},$$

i.e.,

$$\boldsymbol{v}_1 = \boldsymbol{v}_c + \frac{m_1 T_2 \boldsymbol{v}_{dz1} + m_2 T_1 \boldsymbol{v}_{dz2}}{m_1 T_2 + m_2 T_1} + \frac{m_2 T_1}{m_1 T_2 + m_2 T_1}\boldsymbol{v},$$

$$\boldsymbol{v}_2 = \boldsymbol{v}_c + \frac{m_1 T_2 \boldsymbol{v}_{dz1} + m_2 T_1 \boldsymbol{v}_{dz2}}{m_1 T_2 + m_2 T_1} - \frac{m_1 T_2}{m_1 T_2 + m_2 T_1}\boldsymbol{v},$$

gives

$$d\boldsymbol{v}_1 d\boldsymbol{v}_2 = d\boldsymbol{v} d\boldsymbol{v}_c,$$

where we have used the Jacobian

$$J = \frac{d\boldsymbol{v}_1 d\boldsymbol{v}_2}{d\boldsymbol{v} d\boldsymbol{v}_c} = \frac{\partial(v_{1x}, v_{1y}, v_{1z}, v_{2x}, v_{2y}, v_{2z})}{\partial(v_x, v_y, v_z, v_{cx}, v_{cy}, v_{cz})} = \begin{vmatrix} a & & & 1 & & \\ & a & & & 1 & \\ & & a & & & 1 \\ b & & & 1 & & \\ & b & & & 1 & \\ & & b & & & 1 \end{vmatrix} = 1,$$

$$a = \frac{m_2 T_1}{m_1 T_2 + m_2 T_1}, b = -\frac{m_1 T_2}{m_1 T_2 + m_2 T_1},$$

$$v_{dz} \equiv \frac{m_1 T_2 v_{dz1} + m_2 T_1 v_{dz2}}{m_1 T_2 + m_2 T_1}, v_{d0} \equiv v_{dz2} - v_{dz1}.$$

We have

$$\langle \sigma v \rangle_{DM} = \iint d\boldsymbol{v}_1 d\boldsymbol{v}_2 \sigma(v) v f_1(\boldsymbol{v}_1) f_2(\boldsymbol{v}_2) =$$

$$= \left(\frac{m_1 m_2}{(2\pi k_B)^2 T_1 T_2}\right)^{\frac{3}{2}} \iint d\boldsymbol{v} d\boldsymbol{v}_c \sigma(v) v \exp\left(-\frac{m_1 v_{\perp 1}^2 + m_1(v_{\|1} - v_{dz1})^2}{2k_B T_1}\right.$$

$$\left. - \frac{m_2 v_{\perp 2}^2 + m_2(v_{\|2} - v_{dz2})^2}{2k_B T_2}\right).$$

And since

$$\frac{m_1 v_{\perp 1}^2 + m_1(v_{\|1} - v_{dz1})^2}{T_1} + \frac{m_2 v_{\perp 2}^2 + m_2(v_{\|2} - v_{dz2})^2}{T_2}$$

$$= \frac{m_1}{T_1}\left[(v_{cx} + av_x)^2 + (v_{cy} + av_y)^2 + (v_{cz} + v_{dz} - v_{dz1} + av_z)^2\right]$$

$$+ \frac{m_2}{T_2}\left[(v_{cx} + bv_x)^2 + (v_{cy} + bv_y)^2 + (v_{cz} + v_{dz} - v_{dz2} + bv_z)^2\right]$$

$$= \frac{m_1}{T_1}\left[(v_{cx} + av_x)^2 + (v_{cy} + av_y)^2 \right.$$

$$\left. + \left(v_{cz} + \frac{m_2 T_1 v_{d0}}{m_1 T_2 + m_2 T_1} + av_z\right)^2\right]$$

$$+ \frac{m_2}{T_2}\left[(v_{cx} + bv_x)^2 + (v_{cy} + bv_y)^2 + \left(v_{cz} - \frac{m_1 T_2 v_{d0}}{m_1 T_2 + m_2 T_1} + bv_z\right)^2\right]$$

$$= \frac{m_1}{T_1}\left[v_{cx}^2 + a^2 v^2 + 2a v_{cx} v_x + v_{cy}^2 + 2a v_{cy} v_y\right.$$

$$\left. + \left(v_{cz} + \frac{m_2 T_1 v_{d0}}{m_1 T_2 + m_2 T_1}\right)^2 + 2a v_z \left(v_{cz} + \frac{m_2 T_1 v_{d0}}{m_1 T_2 + m_2 T_1}\right)\right]$$

$$+ \frac{m_2}{T_2}\left[v_{cx}^2 + b^2 v^2 + 2b v_{cx} v_x + v_{cy}^2 + 2b v_{cy} v_y\right.$$

$$\left. + \left(v_{cz} - \frac{m_1 T_2 v_{d0}}{m_1 T_2 + m_2 T_1}\right)^2 + 2b v_z \left(v_{cz} - \frac{m_1 T_2 v_{d0}}{m_1 T_2 + m_2 T_1}\right)\right]$$

$$= \frac{m_1}{T_1}\left[v_{cx}^2 + a^2 v^2 + v_{cy}^2 + \left(v_{cz} + \frac{m_2 T_1 v_{d0}}{m_1 T_2 + m_2 T_1}\right)^2\right]$$

$$+ \frac{m_2}{T_2}\left[v_{cx}^2 + b^2 v^2 + v_{cy}^2 + \left(v_{cz} - \frac{m_1 T_2 v_{d0}}{m_1 T_2 + m_2 T_1}\right)^2\right]$$

$$+ 2\frac{m_1 m_2}{m_1 T_2 + m_2 T_1} v_z v_{d0},$$

i.e.,

$$\frac{m_1 v_{\perp 1}^2 + m_1(v_{\parallel 1} - v_{dz1})^2}{T_1} + \frac{m_2 v_{\perp 2}^2 + m_2(v_{\parallel 2} - v_{dz2})^2}{T_2}$$

$$= \frac{m_1}{T_1}\left[v_c^2 + a^2 v^2 + \left(\frac{m_2 T_1 v_{d0}}{m_1 T_2 + m_2 T_1}\right)^2\right]$$

$$+ \frac{m_2}{T_2}\left[v_c^2 + b^2 v^2 + \left(\frac{m_1 T_2 v_{d0}}{m_1 T_2 + m_2 T_1}\right)^2\right] + 2\frac{m_1 m_2}{m_1 T_2 + m_2 T_1} v_z v_{d0}$$

$$= \left(\frac{m_1}{T_1} + \frac{m_2}{T_2}\right) v_c^2 + \frac{m_r}{T_r} v^2 + \frac{m_1 m_2}{m_1 T_2 + m_2 T_1} v_{d0}^2$$

$$+ 2\frac{m_1 m_2}{m_1 T_2 + m_2 T_1} v_z v_{d0}$$

$$= \left(\frac{m_1}{T_1} + \frac{m_2}{T_2}\right) v_c^2 + \frac{m_r}{T_r} v^2 + \frac{m_r}{T_r} v_{d0}^2 + 2\frac{m_r}{T_r} v_z v_{d0},$$

$$v_x = v\sin\theta\cos\phi, v_y = v\sin\theta\sin\phi, v_z = v\cos\theta. \quad d\boldsymbol{v} = v^2 \sin\theta dv d\theta d\phi,$$

we can have

$$\langle\sigma v\rangle_{DM} = \left(\frac{1}{2\pi k_B}\left(\frac{m_1}{T_1}\right.\right.$$

$$\left.\left.+ \frac{m_2}{T_2}\right)\right)^{\frac{3}{2}} \left(\frac{m_r}{2\pi k_B T_r}\right)^{\frac{3}{2}} \iint d\boldsymbol{v} d\boldsymbol{v}_c \sigma(v) v \exp\left(-\frac{1}{2k_B}\left(\frac{m_1}{T_1} + \frac{m_2}{T_2}\right) v_c^2\right.$$

$$\left.- \frac{1}{2k_B}\frac{m_r}{T_r} v^2 - \frac{1}{2k_B}\frac{m_r}{T_r} v_{d0}^2 - \frac{1}{2k_B} 2\frac{m_r}{T_r} v\cos\theta v_{d0}\right)$$

$$= 2\pi \left(\frac{m_r}{2\pi k_B T_r}\right)^{\frac{3}{2}} \iint v^2 \sin\theta dv d\theta \sigma(v) v \exp\left(-\frac{m_r}{2k_B T_r}(v^2 + v_{d0}^2)\right.$$

$$\left.- \frac{m_r}{k_B T_r} v\cos\theta v_{d0}\right)$$

$$= \frac{2k_B T_r}{m_r v_{d0}} 2\pi \left(\frac{m_r}{2\pi k_B T_r}\right)^{\frac{3}{2}} \int_0^\infty v^2 \sigma(v) \exp\left[-\frac{m_r}{2k_B T_r}(v^2 \right.$$

$$\left.+ v_{d0}^2)\right] \sinh\left(\frac{m_r v v_{d0}}{k_B T_r}\right) dv$$

$$= \frac{1}{v_{tr} v_{d0}} \frac{2}{\sqrt{\pi}} \int_0^\infty v^2 \sigma(v) \exp\left[-\frac{(v^2 + v_{d0}^2)}{v_{tr}^2}\right] \sinh\left(2\frac{v v_{d0}}{v_{tr}^2}\right) dv.$$

Here we have used

$$\int_0^\pi \sin\theta \exp\left(-\frac{m_r v v_{d0}}{k_B T_r}\cos\theta\right) d\theta = \left.\frac{\exp\left(-\frac{m_r v v_{d0}}{k_B T_r}\cos\theta\right)}{\frac{m_r v v_{d0}}{k_B T_r}}\right|_0^\pi$$

$$= \frac{\exp\left(\frac{m_r v v_{d0}}{k_B T_r}\right) - \exp\left(-\frac{m_r v v_{d0}}{k_B T_r}\right)}{\frac{m_r v v_{d0}}{k_B T_r}} = \frac{2 k_B T_r}{m_r v v_{d0}} \sinh\left(\frac{m_r v v_{d0}}{k_B T_r}\right).$$

The above results can reduce to Morse2018, Miley1975 and Miley1974.

Note

$$\sinh(x) = \frac{e^x - e^{-x}}{2} \simeq x + \frac{x^3}{6} + \cdots.$$

### 3. Drift Maxwellian at arbitrary directions

For drift Maxwellian at arbitrary directions

$$f_j(v_j) = \left(\frac{m_j}{2\pi k_B T_j}\right)^{\frac{3}{2}} \exp\left[-\frac{m_j(\boldsymbol{v}_j - \boldsymbol{v}_{dj})^2}{2 k_B T_j}\right],$$

we define

$$\boldsymbol{v} = \boldsymbol{v}_1 - \boldsymbol{v}_2, \quad \boldsymbol{v}_c = \frac{m_1 T_2(\boldsymbol{v}_1 - \boldsymbol{v}_{d1}) + m_2 T_1(\boldsymbol{v}_2 - \boldsymbol{v}_{d2})}{m_1 T_2 + m_2 T_1},$$

$$m_r \equiv \frac{m_1 m_2}{m_1 + m_2}, \quad E \equiv \frac{1}{2} m_r v^2,$$

$$T_r \equiv \frac{m_2 T_1 + m_1 T_2}{m_1 + m_2}, \quad \frac{m_1 m_2}{T_1 T_2} = \left(\frac{m_1}{T_1} + \frac{m_2}{T_2}\right)\frac{m_r}{T_r},$$

$$v_{tr} \equiv \sqrt{\frac{2 k_B T_r}{m_r}}.$$

i.e.,

$$\boldsymbol{v}_1 = \boldsymbol{v}_c + \frac{m_1 T_2 \boldsymbol{v}_{d1} + m_2 T_1 \boldsymbol{v}_{d2}}{m_1 T_2 + m_2 T_1} + \frac{m_2 T_1}{m_1 T_2 + m_2 T_1}\boldsymbol{v},$$

$$\boldsymbol{v}_2 = \boldsymbol{v}_c + \frac{m_1 T_2 \boldsymbol{v}_{d1} + m_2 T_1 \boldsymbol{v}_{d2}}{m_1 T_2 + m_2 T_1} - \frac{m_1 T_2}{m_1 T_2 + m_2 T_1}\boldsymbol{v},$$

gives

$$d\boldsymbol{v}_1 d\boldsymbol{v}_2 = d\boldsymbol{v} d\boldsymbol{v}_c,$$

where we have used the Jacobian

$$J = \frac{d\boldsymbol{v}_1 d\boldsymbol{v}_2}{d\boldsymbol{v} d\boldsymbol{v}_c} = \frac{\partial(v_{1x}, v_{1y}, v_{1z}, v_{2x}, v_{2y}, v_{2z})}{\partial(v_x, v_y, v_z, v_{cx}, v_{cy}, v_{cz})} = \begin{vmatrix} a & & & 1 & & \\ & a & & & 1 & \\ & & a & & & 1 \\ b & & & 1 & & \\ & b & & & 1 & \\ & & b & & & 1 \end{vmatrix} = 1,$$

$$a = \frac{m_2 T_1}{m_1 T_2 + m_2 T_1}, b = -\frac{m_1 T_2}{m_1 T_2 + m_2 T_1},$$

$$\boldsymbol{v}_d \equiv \frac{m_1 T_2 \boldsymbol{v}_{d1} + m_2 T_1 \boldsymbol{v}_{d2}}{m_1 T_2 + m_2 T_1}, \boldsymbol{v}_{d0} \equiv \boldsymbol{v}_{d2} - \boldsymbol{v}_{d1}.$$

gives

$$\langle \sigma v \rangle_{DM} = \iint d\boldsymbol{v}_1 d\boldsymbol{v}_2 \sigma(v) v f_1(\boldsymbol{v}_1) f_2(\boldsymbol{v}_2) =$$

$$= \left( \frac{m_1 m_2}{(2\pi k_B)^2 T_1 T_2} \right)^{\frac{3}{2}} \iint d\boldsymbol{v} d\boldsymbol{v}_c \sigma(v) v \exp\left[ -\frac{m_1 (\boldsymbol{v}_1 - \boldsymbol{v}_{d1})^2}{2 k_B T_1} - \frac{m_2 (\boldsymbol{v}_2 - \boldsymbol{v}_{d2})^2}{2 k_B T_2} \right].$$

Since

$$\frac{m_1(\boldsymbol{v}_1 - \boldsymbol{v}_{d1})^2}{T_1} + \frac{m_2(\boldsymbol{v}_2 - \boldsymbol{v}_{d2})^2}{T_2}$$

$$= \frac{m_1}{T_1}\left[(v_{cx} + v_{dx} - v_{dx1} + av_x)^2 + (v_{cy} + v_{dy} - v_{dy1} + av_y)^2\right.$$

$$+ (v_{cz} + v_{dz} - v_{dz1} + av_z)^2\Big]$$

$$+ \frac{m_2}{T_2}\left[(v_{cx} + v_{dx} - v_{dy2} + bv_x)^2 + (v_{cy} + v_{dy} - v_{dy2} + bv_y)^2\right.$$

$$+ (v_{cz} + v_{dz} - v_{dz2} + bv_z)^2\Big]$$

$$= \frac{m_1}{T_1}\left[\left(v_{cx} + \frac{m_2 T_1 v_{d0x}}{m_1 T_2 + m_2 T_1} + av_x\right)^2\right.$$

$$\left. + \left(v_{cy} + \frac{m_2 T_1 v_{d0y}}{m_1 T_2 + m_2 T_1} + av_y\right)^2 + \left(v_{cz} + \frac{m_2 T_1 v_{d0z}}{m_1 T_2 + m_2 T_1} + av_z\right)^2\right]$$

$$+ \frac{m_2}{T_2}\left[\left(v_{cx} - \frac{m_1 T_2 v_{d0x}}{m_1 T_2 + m_2 T_1} + bv_x\right)^2\right.$$

$$\left. + \left(v_{cy} - \frac{m_1 T_2 v_{d0y}}{m_1 T_2 + m_2 T_1} + bv_y\right)^2 + \left(v_{cz} - \frac{m_1 T_2 v_{d0z}}{m_1 T_2 + m_2 T_1} + bv_z\right)^2\right]$$

Where in $x$ direction

$$\frac{m_1}{T_1}\left[\left(v_{cx} + \frac{m_2 T_1 v_{d0x}}{m_1 T_2 + m_2 T_1} + av_x\right)^2\right] + \frac{m_2}{T_2}\left[\left(v_{cx} - \frac{m_1 T_2 v_{d0x}}{m_1 T_2 + m_2 T_1} + bv_x\right)^2\right]$$

$$= \frac{m_1}{T_1}\left[\left(v_{cx} + \frac{m_2 T_1 v_{d0x}}{m_1 T_2 + m_2 T_1}\right)^2 + 2av_x\left(v_{cx} + \frac{m_2 T_1 v_{d0x}}{m_1 T_2 + m_2 T_1}\right)\right.$$

$$\left. + a^2 v_x^2\right]$$

$$+ \frac{m_2}{T_2}\left[\left(v_{cx} - \frac{m_1 T_2 v_{d0x}}{m_1 T_2 + m_2 T_1}\right)^2 + 2bv_x\left(v_{cx} - \frac{m_1 T_2 v_{d0x}}{m_1 T_2 + m_2 T_1}\right) + b^2 v_x^2\right]$$

$$= \frac{m_1}{T_1}\left[v_{cx}^2 + \left(\frac{m_2 T_1 v_{d0x}}{m_1 T_2 + m_2 T_1}\right)^2\right] + \frac{m_2}{T_2}\left[v_{cx}^2 + \left(\frac{m_1 T_2 v_{d0x}}{m_1 T_2 + m_2 T_1}\right)^2\right]$$

$$+ \frac{m_r}{T_r}v_x^2 + 2\frac{m_r}{T_r}v_x v_{d0x}$$

$$= \left(\frac{m_1}{T_1} + \frac{m_2}{T_2}\right)v_{cx}^2 + \frac{m_r}{T_r}v_x^2 + \frac{m_r}{T_r}v_{d0x}^2 + 2\frac{m_r}{T_r}v_x v_{d0x}$$

Also for $y, z$ directions, thus

$$\frac{m_1(\boldsymbol{v}_1 - \boldsymbol{v}_{d1})^2}{T_1} + \frac{m_2(\boldsymbol{v}_2 - \boldsymbol{v}_{d2})^2}{T_2}$$

$$= \left(\frac{m_1}{T_1} + \frac{m_2}{T_2}\right)v_c^2 + \frac{m_r}{T_r}v^2 + \frac{m_r}{T_r}v_{d0}^2 + 2\frac{m_r}{T_r}(v_x v_{d0x} + v_y v_{d0y} + v_z v_{d0z})$$

$$v_x = v\sin\theta\cos\phi, \ v_y = v\sin\theta\sin\phi, v_z = v\cos\theta. \quad d\boldsymbol{v} = v^2 \sin\theta dv d\theta d\phi.$$

So, we have

$$\langle \sigma v \rangle_{DM} = \left(\frac{1}{2\pi k_B}\left(\frac{m_1}{T_1}\right.\right.$$

$$\left.\left.+ \frac{m_2}{T_2}\right)\right)^{\frac{3}{2}} \left(\frac{m_r}{2\pi k_B T_r}\right)^{\frac{3}{2}} \iint d\boldsymbol{v} d\boldsymbol{v}_c \sigma(v) v \exp\left(-\frac{1}{2k_B}\left(\frac{m_1}{T_1} + \frac{m_2}{T_2}\right)v_c^2\right.$$

$$\left. - \frac{1}{2k_B}\frac{m_r}{T_r}v^2 - \frac{1}{2k_B}\frac{m_r}{T_r}v_{d0}^2 - \frac{1}{2k_B}2\frac{m_r}{T_r}(v_x v_{d0x} + v_y v_{d0y} + v_z v_{d0z})\right)$$

$$= \left(\frac{m_r}{2\pi k_B T_r}\right)^{\frac{3}{2}} \int d\boldsymbol{v}\sigma(v) v \exp\left(-\frac{m_r}{2k_B T_r}(v^2 + v_{d0}^2) - \frac{m_r}{k_B T_r}\boldsymbol{v}\cdot\boldsymbol{v}_{d0}\right)$$

$$= \left(\frac{m_r}{2\pi k_B T_r}\right)^{\frac{3}{2}} \iiint v^2 \sin\theta dv d\theta d\phi \sigma(v) v \exp\left[-\frac{m_r}{2k_B T_r}(v^2 + v_{d0}^2)\right.$$

$$\left. - \frac{m_r v}{k_B T_r}(\sin\theta\cos\phi v_{d0x} + \sin\theta\sin\phi v_{d0y} + \cos\theta v_{d0z})\right].$$

Use coordinates rotation, to keep $\boldsymbol{v}_{d0}$ at $z'$ direction, and $v_{d0x'} = v_{d0y'} = 0$. Define new coordinates, gives

$$\langle \sigma v \rangle_{DM} = \left(\frac{m_r}{2\pi k_B T_r}\right)^{\frac{3}{2}} \int d\boldsymbol{v}\sigma(v) v \exp\left(-\frac{m_r}{2k_B T_r}(v^2 + v_{d0}^2) - \frac{m_r}{k_B T_r}\boldsymbol{v}\cdot\boldsymbol{v}_{d0}\right)$$

$$= \left(\frac{m_r}{2\pi k_B T_r}\right)^{\frac{3}{2}} \iiint v^2\sin\theta' dv d\theta' d\phi \sigma(v) v \exp\left[-\frac{m_r}{2k_B T_r}(v^2 + v_{d0}^2)\right.$$

$$\left. - \frac{m_r v}{k_B T_r}(\cos\theta' v_{d0z'})\right]$$

$$= \frac{1}{v_{tr} v_{d0}}\frac{2}{\sqrt{\pi}}\int_0^\infty v^2 \sigma(v)\exp\left[-\frac{(v^2 + v_{d0}^2)}{v_{tr}^2}\right]\sinh\left(2\frac{vv_{d0}}{v_{tr}^2}\right)dv.$$

This result agrees with the previous parallel drift beam case. However, the above result is more general.

Define

$$E_d = k_B T_d = \frac{m_r v_{d0}^2}{2},$$

and

$$dE = m_r v dv, \quad v = \sqrt{\frac{2E}{m_r}}, \quad v_{d0} = \sqrt{\frac{2E_d}{m_r}},$$

gives

$$\langle \sigma v \rangle_{DM} = \frac{1}{v_{tr} v_{d0}} \frac{2}{\sqrt{\pi}} \int_0^\infty v^2 \sigma(v) \exp\left[-\frac{(v^2 + v_{d0}^2)}{v_{tr}^2}\right] \sinh\left(2\frac{v v_{d0}}{v_{tr}^2}\right) dv$$

$$= \sqrt{\frac{2}{\pi m_r k_B T_r k_B T_d}} \int_0^\infty \sqrt{E} \sigma(E) \exp\left[-\frac{E + E_d}{k_B T_r}\right] \sinh\left(\frac{2\sqrt{E E_d}}{k_B T_r}\right) dE.$$

Note $v_d \to 0, E_d \to 0$,

$$\sinh\left(\frac{2\sqrt{E E_d}}{k_B T_r}\right) \to \frac{2\sqrt{E E_d}}{k_B T_r},$$

thus

$$\langle \sigma v \rangle_{DM} \to \sqrt{\frac{8}{\pi m_r (k_B T_r)^3}} \int_0^\infty E \sigma(E) \exp\left[-\frac{E}{k_B T_r}\right] dE,$$

which agrees with the previous Maxwellian case.

Note also for cold beam-target $v_{tr} \to 0, T_r \to 0$,

$$\langle \sigma v \rangle_{DM} \to \frac{1}{v_{tr} v_{d0}} \frac{1}{\sqrt{\pi}} \int_0^\infty v^2 \sigma(v) \exp\left[-\frac{(v - v_{d0})^2}{v_{tr}^2}\right] dv \to v_{d0} \sigma(v_{d0}),$$

where we have used the delta function

$$\delta(x) = \lim_{\epsilon \to 0^+} \frac{1}{\epsilon \sqrt{\pi}} e^{-\frac{x^2}{\epsilon^2}}.$$

Also, if $T_1 \to 0, T_2 \neq 0$, i.e., cold beam to hot plasma, thus

$$T_r = \frac{m_2 T_1 + m_1 T_2}{m_1 + m_2} \to \frac{m_1 T_2}{m_1 + m_2},$$

$$v_{tr} = \sqrt{\frac{2 k_B T_r}{m_r}} \to \sqrt{\frac{2 k_B T_2}{m_2}}$$

which only changes the effective temperature $T_r$, and does not change the

integral form.

## 4. Bi-Maxwellian distributions

For bi-Maxwellian distributions

$$f_j(\boldsymbol{v}_j) = \left(\frac{m_j}{2\pi k_B T_{\|j}}\right)^{\frac{1}{2}} \left(\frac{m_j}{2\pi k_B T_{\perp j}}\right)^{1} \exp\left(-\frac{m_j v_{\perp j}^2}{2k_B T_{\perp j}} - \frac{m_j v_{\|j}^2}{2k_B T_{\|j}}\right),$$

$$v_{\perp j}^2 = v_{xj}^2 + v_{yj}^2, \qquad v_{\|j} = v_{zj},$$

$$\int d\boldsymbol{v}\, f_j(\boldsymbol{v}_j) = 1,$$

Define

$$T_{\|j} = (1+\delta_j)T_j, \qquad T_{\perp j} = \left(1 - \frac{\delta_j}{2}\right)T_j.$$

and

$$\boldsymbol{v} = \boldsymbol{v}_1 - \boldsymbol{v}_2, \qquad m_r \equiv \frac{m_1 m_2}{m_1 + m_2},$$

$$T_{\perp r} \equiv \frac{m_2 T_{\perp 1} + m_1 T_{\perp 2}}{m_1 + m_2}, \qquad T_{\|r} \equiv \frac{m_2 T_{\|1} + m_1 T_{\|2}}{m_1 + m_2},$$

and

$$T_r \equiv \frac{m_2 T_1 + m_1 T_2}{m_1 + m_2}, \qquad \delta \equiv \frac{m_2 T_1 \delta_1 + m_1 T_2 \delta_2}{m_1 T_2 + m_2 T_1}$$

thus

$$T_{\|r} = (1+\delta)T_r, \qquad T_{\perp r} = \left(1 - \frac{\delta}{2}\right)T_r.$$

And define

$$\boldsymbol{v}_c \equiv \frac{m_2 T_{\perp 1} \boldsymbol{v}_{\perp 2} + m_1 T_{\perp 2} \boldsymbol{v}_{\perp 1}}{m_1 T_{\perp 2} + m_2 T_{\perp 1}} + \frac{m_2 T_{\|1} \boldsymbol{v}_{\|2} + m_1 T_{\|2} \boldsymbol{v}_{\|1}}{m_1 T_{\|2} + m_2 T_{\|1}},$$

i.e.,

$$\boldsymbol{v}_{\perp 1} = \boldsymbol{v}_{c\perp} + \frac{m_2 T_{\perp 1}}{m_1 T_{\perp 2} + m_2 T_{\perp 1}} \boldsymbol{v}_\perp, \qquad \boldsymbol{v}_{\perp 2} = \boldsymbol{v}_{c\perp} - \frac{m_1 T_{\perp 2}}{m_1 T_{\perp 2} + m_2 T_{\perp 1}} \boldsymbol{v}_\perp,$$

$$\boldsymbol{v}_{\|1} = \boldsymbol{v}_{c\|} + \frac{m_2 T_{\|1}}{m_1 T_{\|2} + m_2 T_{\|1}} \boldsymbol{v}_\|, \qquad \boldsymbol{v}_{\|2} = \boldsymbol{v}_{c\|} - \frac{m_1 T_{\|2}}{m_1 T_{\|2} + m_2 T_{\|1}} \boldsymbol{v}_\|.$$

gives

$$d\boldsymbol{v}_1 d\boldsymbol{v}_2 = d\boldsymbol{v} d\boldsymbol{v}_c,$$

where we have used the below Jacobian

$$J = \frac{d\boldsymbol{v}_1 d\boldsymbol{v}_2}{d\boldsymbol{v} d\boldsymbol{v}_c} = \frac{\partial(v_{1x}, v_{1y}, v_{1z}, v_{2x}, v_{2y}, v_{2z})}{\partial(v_x, v_y, v_z, v_{cx}, v_{cy}, v_{cz})} = \begin{vmatrix} a_\perp & & & 1 & & \\ & a_\perp & & & 1 & \\ & & a_\parallel & & & 1 \\ b_\perp & & & 1 & & \\ & b_\perp & & & 1 & \\ & & b_\parallel & & & 1 \end{vmatrix} = 1,$$

$$a_\perp = \frac{m_2 T_{\perp 1}}{m_1 T_{\perp 2} + m_2 T_{\perp 1}}, \quad b_\perp = -\frac{m_1 T_{\perp 2}}{m_1 T_{\perp 2} + m_2 T_{\perp 1}},$$

$$a_\parallel = \frac{m_2 T_{\parallel 1}}{m_1 T_{\parallel 2} + m_2 T_{\parallel 1}}, \quad b_\parallel = -\frac{m_1 T_{\parallel 2}}{m_1 T_{\parallel 2} + m_2 T_{\parallel 1}}.$$

We have

$$\frac{m_1 v_{\parallel 1}^2}{T_{\parallel 1}} + \frac{m_2 v_{\parallel 2}^2}{T_{\parallel 2}} = \frac{m_1}{T_{\parallel 1}}(v_{c\parallel} + a_\parallel v_\parallel)^2 + \frac{m_2}{T_{\parallel 2}}(v_{c\parallel} + b_\parallel v_\parallel)^2$$

$$= \left(\frac{m_1}{T_{\parallel 1}} + \frac{m_2}{T_{\parallel 2}}\right) v_{c\parallel}^2 + 2a_\parallel \frac{m_1}{T_{\parallel 1}} v_{c\parallel} v_\parallel + 2b_\parallel \frac{m_2}{T_{\parallel 2}} v_{c\parallel} v_\parallel + \frac{m_1}{T_{\parallel 1}} a_\parallel^2 v_\parallel^2$$

$$+ \frac{m_2}{T_{\parallel 2}} b_\parallel^2 \; v_\parallel^2$$

$$= \left(\frac{m_1}{T_{\parallel 1}} + \frac{m_2}{T_{\parallel 2}}\right) v_{c\parallel}^2$$

$$+ \left[\frac{m_1}{T_{\parallel 1}}\left(\frac{m_2 T_{\parallel 1}}{m_1 T_{\parallel 2} + m_2 T_{\parallel 1}}\right)^2 + \frac{m_2}{T_{\parallel 2}}\left(\frac{m_1 T_{\parallel 2}}{m_1 T_{\parallel 2} + m_2 T_{\parallel 1}}\right)^2\right] v_\parallel^2$$

$$= \left(\frac{m_1}{T_{\parallel 1}} + \frac{m_2}{T_{\parallel 2}}\right) v_{c\parallel}^2 + \frac{m_r}{T_{\parallel r}} v_\parallel^2.$$

also

$$\frac{m_1 v_{\perp 1}^2}{T_{\perp 1}} + \frac{m_2 v_{\perp 2}^2}{T_{\perp 2}} = \left(\frac{m_1}{T_{\perp 1}} + \frac{m_2}{T_{\perp 2}}\right) v_{c\perp}^2 + \frac{m_r}{T_{\perp r}} v_\perp^2.$$

Thus

$$\langle \sigma v \rangle = \iint d\boldsymbol{v}_1 d\boldsymbol{v}_2 \sigma(v) v f_1(\boldsymbol{v}_1) f_2(\boldsymbol{v}_2)$$

$$= \left(\frac{m_1}{2\pi k_B T_{\|1}}\right)^{\frac{1}{2}} \left(\frac{m_1}{2\pi k_B T_{\perp 1}}\right)^{1} \left(\frac{m_2}{2\pi k_B T_{\|2}}\right)^{\frac{1}{2}} \left(\frac{m_2}{2\pi k_B T_{\perp 2}}\right)^{1} \iint d\boldsymbol{v}_1 d\boldsymbol{v}_2 \sigma(v) v \exp\left(-\frac{m_1 v_{\perp 1}^2}{2k_B T_{\perp 1}}\right.$$
$$\left. -\frac{m_1 v_{\|1}^2}{2k_B T_{\|1}} - \frac{m_2 v_{\perp 2}^2}{2k_B T_{\perp 2}} - \frac{m_2 v_{\|2}^2}{2k_B T_{\|2}}\right)$$

$$= \left(\frac{m_1}{2\pi k_B T_{\|1}} \frac{m_2}{2\pi k_B T_{\|2}}\right)^{\frac{1}{2}} \left(\frac{m_1}{2\pi k_B T_{\perp 1}} \frac{m_2}{2\pi k_B T_{\perp 2}}\right)^{1} \iint d\boldsymbol{v} d\boldsymbol{v}_c \sigma(v) v \exp\left\{-\frac{1}{2k_B}\left[\left(\frac{m_1}{T_{\|1}}\right.\right.\right.$$
$$\left.\left.\left. + \frac{m_2}{T_{\|2}}\right) v_{c\|}^2 + \frac{m_r}{T_{\|r}} v_{\|}^2 + \left(\frac{m_1}{T_{\perp 1}} + \frac{m_2}{T_{\perp 2}}\right) v_{c\perp}^2 + \frac{m_r}{T_{\perp r}} v_{\perp}^2\right]\right\}$$

$$= \left(\frac{m_r}{2\pi k_B}\right)^{\frac{3}{2}} \frac{1}{T_{\|r}^{1/2} T_{\perp r}} \int d\boldsymbol{v} \sigma(v) v \exp\left\{-\frac{1}{2k_B}\left[\frac{m_r}{T_{\|r}} v_{\|}^2 + \frac{m_r}{T_{\perp r}} v_{\perp}^2\right]\right\}.$$

Note we have used

$$\frac{m_1 m_2}{T_1 T_2} = \left(\frac{m_1}{T_1} + \frac{m_2}{T_2}\right) \frac{m_r}{T_r}.$$

The above result agrees with Kolmes21 Eq.(12).

Note also

$$E = E_\perp + E_\|, E_\| = \frac{1}{2} m_r v_\|^2, E_\perp = \frac{1}{2} m_r v_\perp^2,$$

$$v_x = \sqrt{2E_\perp/m_r}\cos\phi, v_y = \sqrt{2E_\perp/m_r}\sin\phi, v_z = \pm\sqrt{2E_\|/m_r}.$$
$$\phi = [0, 2\pi], E_\perp = [0, \infty), E_\| = [0, \infty).$$

$$\left| \begin{array}{ccc} 0 & 0 & \frac{1}{2}\sqrt{\frac{2}{E_\| m_r}} \\ \frac{1}{2}\sqrt{\frac{2}{E_\perp m_r}}\cos\phi & \frac{1}{2}\sqrt{\frac{2}{E_\perp m_r}}\sin\phi & 0 \\ -\sqrt{2E_\perp/m_r}\sin\phi & \sqrt{2E_\perp/m_r}\cos\phi & 0 \end{array} \right| = \frac{1}{2m_r}\sqrt{\frac{2}{E_\| m_r}},$$

gives

$$d\boldsymbol{v} = \frac{1}{2m_r}\sqrt{\frac{2}{E_\| m_r}} dE_\perp dE_\| d\phi.$$

We have

$$\langle \sigma v \rangle = \left(\frac{m_r}{2\pi k_B}\right)^{\frac{3}{2}} \frac{1}{T_{\|r}^{1/2} T_{\perp r}} \int d\boldsymbol{v}\, \sigma(v) v \exp\left\{-\frac{1}{2k_B}\left[\frac{m_r}{T_{\|r}} v_\|^2 + \frac{m_r}{T_{\perp r}} v_\perp^2\right]\right\}$$

$$= \left(\frac{m_r}{2\pi k_B}\right)^{\frac{3}{2}} \frac{1}{T_{\|r}^{1/2} T_{\perp r}} \int d\boldsymbol{v}\, \sigma(v) v \exp\left\{-\frac{E_\|}{k_B T_{\|r}} - \frac{E_\perp}{k_B T_{\perp r}}\right\}$$

$$= \left(\frac{m_r}{2\pi k_B}\right)^{\frac{3}{2}} \frac{1}{T_{\|r}^{1/2} T_{\perp r}} \int_0^\infty \int_0^\infty 2$$

$$\cdot 2\pi \frac{1}{2m_r} \sqrt{\frac{2}{E_\| m_r}}\, dE_\perp dE_\|\, \sigma(v) \sqrt{\frac{2E}{m_r}} \exp\left\{-\frac{E_\|}{k_B T_{\|r}} - \frac{E_\perp}{k_B T_{\perp r}}\right\}$$

$$= \left(\frac{1}{\pi k_B}\right)^{\frac{3}{2}} \frac{1}{T_{\|r}^{1/2} T_{\perp r}} \int_0^\infty \int_0^\infty \pi dE_\perp dE_\|\, \sigma(E) \sqrt{\frac{2E}{m_r E_\|}} \exp\left\{-\frac{E_\|}{k_B T_{\|r}} - \frac{E_\perp}{k_B T_{\perp r}}\right\}$$

$$= \left(\frac{2}{\pi m_r k_B^3}\right)^{\frac{1}{2}} \frac{1}{T_{\|r}^{1/2} T_{\perp r}} \int_0^\infty \int_0^\infty dE_\perp dE_\|\, \sigma(E) \sqrt{\frac{E}{E_\|}} \exp\left\{-\left[\frac{E_\|}{k_B T_{\|r}} + \frac{E_\perp}{k_B T_{\perp r}}\right]\right\},$$

which agrees with Kolmes21 Eq.(13).

For $T_{\perp r} - T_{\|r} \geq 0$, define

$$E = E_\perp + E_\|, \qquad t^2 = \frac{E_\|}{k_B T_{\|r}} - \frac{E_\|}{k_B T_{\perp r}} = E_\| \frac{(T_{\perp r} - T_{\|r})}{k_B T_{\|r} T_{\perp r}},$$

i.e.,

$$E_\| = t^2 \frac{k_B T_{\|r} T_{\perp r}}{(T_{\perp r} - T_{\|r})}, \qquad E_\perp = E - t^2 \frac{k_B T_{\|r} T_{\perp r}}{(T_{\perp r} - T_{\|r})},$$

hence

$$dE_\| = 2t \frac{k_B T_{\|r} T_{\perp r}}{(T_{\perp r} - T_{\|r})} dt, \qquad dE_\perp = dE - 2t \frac{k_B T_{\|r} T_{\perp r}}{(T_{\perp r} - T_{\|r})} dt,$$

$$dE_\| dE_\perp = 2t \frac{k_B T_{\|r} T_{\perp r}}{(T_{\perp r} - T_{\|r})} dE dt.$$

We have

$$\langle \sigma v \rangle = \left(\frac{2}{\pi m_r k_B^3}\right)^{\frac{1}{2}} \frac{1}{T_{\|r}^{1/2} T_{\perp r}} \int_0^\infty \int_0^\infty dE_\perp dE_\| \, \sigma(E) \sqrt{\frac{E}{E_\|}} \exp\left\{-\left[\frac{E_\|}{k_B T_{\|r}} + \frac{E_\perp}{k_B T_{\perp r}}\right]\right\}$$

$$= \left(\frac{2}{\pi m_r k_B^3}\right)^{\frac{1}{2}} \frac{1}{T_{\|r}^{1/2} T_{\perp r}} \int_0^\infty dE \int_0^{\sqrt{E\frac{(T_{\perp r} - T_{\|r})}{k_B T_{\|r} T_{\perp r}}}} dt \exp\left\{-\left[\frac{t^2 \frac{k_B T_{\|r} T_{\perp r}}{(T_{\perp r} - T_{\|r})}}{k_B T_{\|r}}\right.\right.$$

$$\left.\left. + \frac{E - t^2 \frac{k_B T_{\|r} T_{\perp r}}{(T_{\perp r} - T_{\|r})}}{k_B T_{\perp r}}\right]\right\} \sigma(E) 2t \frac{k_B T_{\|r} T_{\perp r}}{(T_{\perp r} - T_{\|r})} \sqrt{\frac{E}{t^2 \frac{k_B T_{\|r} T_{\perp r}}{(T_{\perp r} - T_{\|r})}}}$$

$$= \left(\frac{8}{\pi m_r k_B^2 T_{\perp r}(T_{\perp r} - T_{\|r})}\right)^{\frac{1}{2}} \int_0^\infty dE \int_0^{\sqrt{E\frac{(T_{\perp r} - T_{\|r})}{k_B T_{\|r} T_{\perp r}}}} dt \, \sigma(E) \sqrt{E} \exp\left\{-\left[t^2 + \frac{E}{k_B T_{\perp r}}\right]\right\}$$

$$= \left(\frac{2}{m_r k_B^2 T_{\perp r}(T_{\perp r} - T_{\|r})}\right)^{\frac{1}{2}} \int_0^\infty dE \, \sigma(E) \sqrt{E} \exp\left\{-\frac{E}{k_B T_{\perp r}}\right\} \mathrm{erf}\left[\sqrt{E\frac{(T_{\perp r} - T_{\|r})}{k_B T_{\|r} T_{\perp r}}}\right],$$

which agrees with Kolmes21 Eq.(14). Here we have used the error function

$$\mathrm{erf}(z) = \frac{2}{\sqrt{\pi}} \int_0^z e^{-t^2} dt \simeq \frac{2}{\sqrt{\pi}}\left(z - \frac{z^3}{3} + \cdots\right), \qquad \mathrm{erf}(-z) = -\mathrm{erf}(z).$$

Similarly, we can obtain also for $T_{\perp r} - T_{\|r} < 0$, with the use of

$$\mathrm{erf}(iz) = i \cdot \mathrm{erfi}(z), \qquad \mathrm{erfi}(z) = \frac{2}{\sqrt{\pi}} \int_0^z e^{t^2} dt.$$

For $T_{\|r} - T_{\perp r} > 0$, define

$$E = E_\perp + E_\|, \qquad t^2 = \frac{E_\|}{k_B T_{\perp r}} - \frac{E_\|}{k_B T_{\|r}} = E_\| \frac{(T_{\|r} - T_{\perp r})}{k_B T_{\|r} T_{\perp r}},$$

i.e.,

$$E_\| = t^2 \frac{k_B T_{\|r} T_{\perp r}}{(T_{\|r} - T_{\perp r})}, \qquad E_\perp = E - t^2 \frac{k_B T_{\|r} T_{\perp r}}{(T_{\|r} - T_{\perp r})},$$

hence

$$dE_\| = 2t \frac{k_B T_{\|r} T_{\perp r}}{(T_{\|r} - T_{\perp r})} dt, \qquad dE_\perp = dE - 2t \frac{k_B T_{\|r} T_{\perp r}}{(T_{\|r} - T_{\perp r})} dt,$$

$$dE_\| dE_\perp = 2t \frac{k_B T_{\|r} T_{\perp r}}{(T_{\|r} - T_{\perp r})} dE dt.$$

We have

$$\langle \sigma v \rangle = \left(\frac{2}{\pi m_r k_B^3}\right)^{\frac{1}{2}} \frac{1}{T_{\|r}^{1/2} T_{\perp r}} \int_0^\infty \int_0^\infty dE_\perp dE_\| \, \sigma(E) \sqrt{\frac{E}{E_\|}} \exp\left\{-\left[\frac{E_\|}{k_B T_{\|r}} + \frac{E_\perp}{k_B T_{\perp r}}\right]\right\}$$

$$= \left(\frac{2}{\pi m_r k_B^3}\right)^{\frac{1}{2}} \frac{1}{T_{\|r}^{1/2} T_{\perp r}} \int_0^\infty dE \int_0^{\sqrt{E\frac{(T_{\|r}-T_{\perp r})}{k_B T_{\|r} T_{\perp r}}}} dt \exp\left\{-\left[\frac{t^2 \frac{k_B T_{\|r} T_{\perp r}}{(T_{\|r}-T_{\perp r})}}{k_B T_{\|r}}\right.\right.$$

$$\left.\left. + \frac{E - t^2 \frac{k_B T_{\|r} T_{\perp r}}{(T_{\|r}-T_{\perp r})}}{k_B T_{\perp r}}\right]\right\} \sigma(E) 2t \frac{k_B T_{\|r} T_{\perp r}}{(T_{\|r}-T_{\perp r})} \sqrt{\frac{E}{t^2 \frac{k_B T_{\|r} T_{\perp r}}{(T_{\|r}-T_{\perp r})}}}$$

$$= \left(\frac{8}{\pi m_r k_B^2 T_{\perp r} (T_{\|r}-T_{\perp r})}\right)^{\frac{1}{2}} \int_0^\infty dE \int_0^{\sqrt{E\frac{(T_{\|r}-T_{\perp r})}{k_B T_{\|r} T_{\perp r}}}} dt \exp\left\{t^2 - \frac{E}{k_B T_{\perp r}}\right\} \sigma(E) \sqrt{E}$$

$$= \left(\frac{2}{m_r k_B^2 T_{\perp r} (T_{\|r}-T_{\perp r})}\right)^{\frac{1}{2}} \int_0^\infty dE \, \sigma(E) \sqrt{E} \exp\left\{-\frac{E}{k_B T_{\perp r}}\right\} \text{erfi}\left[\sqrt{E\frac{(T_{\|r}-T_{\perp r})}{k_B T_{\|r} T_{\perp r}}}\right]$$

$$= \left(\frac{2}{m_r k_B^2 T_{\perp r} (T_{\perp r}-T_{\|r})}\right)^{\frac{1}{2}} \int_0^\infty dE \, \sigma(E) \sqrt{E} \exp\left\{-\frac{E}{k_B T_{\perp r}}\right\} \text{erf}\left[\sqrt{E\frac{(T_{\perp r}-T_{\|r})}{k_B T_{\|r} T_{\perp r}}}\right].$$

We see that the final form of the integral is the same as that for $T_{\perp r} - T_{\|r} \geq 0$.

## 5. Parallel drift bi-Maxwellian distributions

For parallel drift bi-Maxwellian distributions

$$f_j(v_j) = \left(\frac{m_j}{2\pi k_B T_{\|j}}\right)^{\frac{1}{2}} \left(\frac{m_j}{2\pi k_B T_{\perp j}}\right)^1 \exp\left(-\frac{m_j v_{\perp j}^2}{2 k_B T_{\perp j}}\right) \exp\left(-\frac{m_j (v_{\|j} - v_{dzj})^2}{2 k_B T_{\|j}}\right),$$

$$\int d\boldsymbol{v}\, f_j(v_j) = 1,$$

$$v_{\perp j}^2 = v_{xj}^2 + v_{yj}^2, \qquad v_{\|j} = v_{zj},$$

we define

$$\boldsymbol{v} = \boldsymbol{v}_1 - \boldsymbol{v}_2,$$

$$m_r \equiv \frac{m_1 m_2}{m_1 + m_2}, \qquad E \equiv \frac{1}{2} m_r v^2,$$

$$T_{\perp r} \equiv \frac{m_2 T_{\perp 1} + m_1 T_{\perp 2}}{m_1 + m_2}, \qquad T_{\|r} \equiv \frac{m_2 T_{\|1} + m_1 T_{\|2}}{m_1 + m_2},$$

$$\boldsymbol{v}_c \equiv \frac{m_2 T_{\perp 1} \boldsymbol{v}_{\perp 2} + m_1 T_{\perp 2} \boldsymbol{v}_{\perp 1}}{m_1 T_{\perp 2} + m_2 T_{\perp 1}} + \frac{m_2 T_{\|1}(\boldsymbol{v}_{\|2} - \boldsymbol{v}_{dz2}) + m_1 T_{\|2}(\boldsymbol{v}_{\|1} - \boldsymbol{v}_{dz1})}{m_1 T_{\|2} + m_2 T_{\|1}}$$

i.e.,

$$\boldsymbol{v}_{\perp 1} = \boldsymbol{v}_{c\perp} + \frac{m_2 T_{\perp 1}}{m_1 T_{\perp 2} + m_2 T_{\perp 1}} \boldsymbol{v}_\perp, \qquad \boldsymbol{v}_{\perp 2} = \boldsymbol{v}_{c\perp} - \frac{m_1 T_{\perp 2}}{m_1 T_{\perp 2} + m_2 T_{\perp 1}} \boldsymbol{v}_\perp,$$

$$\boldsymbol{v}_{\|1} = \boldsymbol{v}_{c\|} + \frac{m_1 T_{\|2} \boldsymbol{v}_{dz1} + m_2 T_{\|1} \boldsymbol{v}_{dz2}}{m_1 T_{\|2} + m_2 T_{\|1}} + \frac{m_2 T_{\|1}}{m_1 T_{\|2} + m_2 T_{\|1}} \boldsymbol{v}_\|,$$

$$\boldsymbol{v}_{\|2} = \boldsymbol{v}_{c\|} + \frac{m_1 T_{\|2} \boldsymbol{v}_{dz1} + m_2 T_{\|1} \boldsymbol{v}_{dz2}}{m_1 T_{\|2} + m_2 T_{\|1}} - \frac{m_1 T_{\|2}}{m_1 T_{\|2} + m_2 T_{\|1}} \boldsymbol{v}_\|.$$

gives

$$d\boldsymbol{v}_1 d\boldsymbol{v}_2 = d\boldsymbol{v}_c d\boldsymbol{v}.$$

where we have used Jacobian

$$J = \frac{d\boldsymbol{v}_1 d\boldsymbol{v}_2}{d\boldsymbol{v} d\boldsymbol{v}_c} = \frac{\partial(v_{1x}, v_{1y}, v_{1z}, v_{2x}, v_{2y}, v_{2z})}{\partial(v_x, v_y, v_z, v_{cx}, v_{cy}, v_{cz})} = \begin{vmatrix} a_\perp & & & 1 & & \\ & a_\perp & & & 1 & \\ & & a_\| & & & 1 \\ b_\perp & & & 1 & & \\ & b_\perp & & & 1 & \\ & & b_\| & & & 1 \end{vmatrix} = 1,$$

$$a_\perp = \frac{m_2 T_{\perp 1}}{m_1 T_{\perp 2} + m_2 T_{\perp 1}}, b_\perp = -\frac{m_1 T_{\perp 2}}{m_1 T_{\perp 2} + m_2 T_{\perp 1}},$$

$$a_\| = \frac{m_2 T_{\|1}}{m_1 T_{\|2} + m_2 T_{\|1}}, b_\| = -\frac{m_1 T_{\|2}}{m_1 T_{\|2} + m_2 T_{\|1}},$$

$$\boldsymbol{v}_{dz} \equiv \frac{m_1 T_{\|2} \boldsymbol{v}_{dz1} + m_2 T_{\|1} \boldsymbol{v}_{dz2}}{m_1 T_{\|2} + m_2 T_{\|1}}, \boldsymbol{v}_{d0} \equiv \boldsymbol{v}_{dz2} - \boldsymbol{v}_{dz1}.$$

gives

$$\frac{m_1 v_{\perp 1}^2}{T_{\perp 1}} + \frac{m_2 v_{\perp 2}^2}{T_{\perp 2}}$$

$$= \frac{m_1}{T_{\perp 1}}\left[\boldsymbol{v}_{c\perp} + \frac{m_2 T_{\perp 1}}{m_1 T_{\perp 2} + m_2 T_{\perp 1}}\boldsymbol{v}_\perp\right]^2$$

$$+ \frac{m_2}{T_{\perp 2}}\left[\boldsymbol{v}_{c\perp} - \frac{m_1 T_{\perp 2}}{m_1 T_{\perp 2} + m_2 T_{\perp 1}}\boldsymbol{v}_\perp\right]^2$$

$$= \left(\frac{m_1}{T_{\perp 1}} + \frac{m_2}{T_{\perp 2}}\right)\boldsymbol{v}_{c\perp}^2 + \frac{m_1 m_2}{m_1 T_{\perp 2} + m_2 T_{\perp 1}}\boldsymbol{v}_\perp^2$$

$$= \left(\frac{m_1}{T_{\perp 1}} + \frac{m_2}{T_{\perp 2}}\right)\boldsymbol{v}_{c\perp}^2 + \frac{m_r}{T_{\perp r}}\boldsymbol{v}_\perp^2,$$

and

$$\frac{m_1(v_{\parallel 1} - v_{dz1})^2}{T_{\parallel 1}} + \frac{m_2(v_{\parallel 2} - v_{dz2})^2}{T_{\parallel 2}}$$

$$= \frac{m_1}{T_{\parallel 1}}\left(\boldsymbol{v}_{c\parallel} + \frac{m_2 T_{\parallel 1}\boldsymbol{v}_{dz2} - m_2 T_{\parallel 1}\boldsymbol{v}_{dz1}}{m_1 T_{\parallel 2} + m_2 T_{\parallel 1}} + \frac{m_2 T_{\parallel 1}}{m_1 T_{\parallel 2} + m_2 T_{\parallel 1}}\boldsymbol{v}_\parallel\right)^2$$

$$+ \frac{m_2}{T_{\parallel 2}}\left(\boldsymbol{v}_{c\parallel} + \frac{m_1 T_{\parallel 2}\boldsymbol{v}_{dz1} - m_1 T_{\parallel 2}\boldsymbol{v}_{dz2}}{m_1 T_{\parallel 2} + m_2 T_{\parallel 1}} - \frac{m_1 T_{\parallel 2}}{m_1 T_{\parallel 2} + m_2 T_{\parallel 1}}\boldsymbol{v}_\parallel\right)^2$$

$$= \frac{m_1}{T_{\parallel 1}}\left(\boldsymbol{v}_{c\parallel} + \frac{m_2 T_{\parallel 1}\boldsymbol{v}_{dz2} - m_2 T_{\parallel 1}\boldsymbol{v}_{dz1}}{m_1 T_{\parallel 2} + m_2 T_{\parallel 1}}\right)^2$$

$$+ \frac{m_2}{T_{\parallel 2}}\left(\boldsymbol{v}_{c\parallel} + \frac{m_1 T_{\parallel 2}\boldsymbol{v}_{dz1} - m_1 T_{\parallel 2}\boldsymbol{v}_{dz2}}{m_1 T_{\parallel 2} + m_2 T_{\parallel 1}}\right)^2 + \frac{m_1 m_2}{m_1 T_{\parallel 2} + m_2 T_{\parallel 1}}\boldsymbol{v}_\parallel^2$$

$$+ 2\frac{m_1 m_2}{m_1 T_{\parallel 2} + m_2 T_{\parallel 1}}\boldsymbol{v}_\parallel v_{d0}$$

$$= \left(\frac{m_1}{T_{\parallel 1}} + \frac{m_2}{T_{\parallel 2}}\right)\boldsymbol{v}_{c\parallel}^2 + \frac{m_1 m_2}{m_1 T_{\parallel 2} + m_2 T_{\parallel 1}}v_{d0}^2 + \frac{m_1 m_2}{m_1 T_{\parallel 2} + m_2 T_{\parallel 1}}\boldsymbol{v}_\parallel^2$$

$$+ 2\frac{m_1 m_2}{m_1 T_{\parallel 2} + m_2 T_{\parallel 1}}\boldsymbol{v}_\parallel v_{d0}$$

$$= \left(\frac{m_1}{T_{\parallel 1}} + \frac{m_2}{T_{\parallel 2}}\right)\boldsymbol{v}_{c\parallel}^2 + \frac{m_r}{T_{\parallel r}}v_{d0}^2 + \frac{m_r}{T_{\parallel r}}\boldsymbol{v}_\parallel^2 + 2\frac{m_r}{T_{\parallel r}}\boldsymbol{v}_\parallel v_{d0},$$

thus

$$\langle \sigma v \rangle = \iint d\boldsymbol{v}_1 d\boldsymbol{v}_2 \sigma(v) v f_1(\boldsymbol{v}_1) f_2(\boldsymbol{v}_2)$$

$$= \left(\frac{m_1}{2\pi k_B T_{\|1}} \frac{m_2}{2\pi k_B T_{\|2}}\right)^{\frac{1}{2}} \left(\frac{m_1}{2\pi k_B T_{\perp 1}} \frac{m_2}{2\pi k_B T_{\perp 2}}\right)^1 \iint d\boldsymbol{v}_1 d\boldsymbol{v}_2 \sigma(v) v \exp\left(-\frac{m_1 v_{\perp 1}^2}{2 k_B T_{\perp 1}}\right.$$
$$\left. -\frac{m_2 v_{\perp 2}^2}{2 k_B T_{\perp 2}} - \frac{m_1(v_{\|1} - v_{dz1})^2}{2 k_B T_{\|1}} - \frac{m_2(v_{\|2} - v_{dz2})^2}{2 k_B T_{\|2}}\right)$$

$$= \left(\frac{m_1}{2\pi k_B T_{\|1}} \frac{m_2}{2\pi k_B T_{\|2}}\right)^{\frac{1}{2}} \left(\frac{m_1}{2\pi k_B T_{\perp 1}} \frac{m_2}{2\pi k_B T_{\perp 2}}\right)^1 \iint d\boldsymbol{v} d\boldsymbol{v}_c \sigma(v) v \exp\left\{-\frac{1}{2 k_B}\left[\left(\frac{m_1}{T_{\perp 1}}\right.\right.\right.$$
$$\left.\left.\left. + \frac{m_2}{T_{\perp 2}}\right) v_{c\perp}^2 + \frac{m_r}{T_{\perp r}} v_\perp^2 + \left(\frac{m_1}{T_{\|1}} + \frac{m_2}{T_{\|2}}\right) v_{c\|}^2 + \frac{m_r}{T_{\|r}} v_{d0}^2 + \frac{m_r}{T_{\|r}} v_\|^2 + 2 \frac{m_r}{T_{\|r}} v_\| v_{d0}\right]\right\}$$

$$= \left(\frac{m_1}{2\pi k_B T_{\|1}} \frac{m_2}{2\pi k_B T_{\|2}}\right)^{\frac{1}{2}} \left(\frac{m_1}{2\pi k_B T_{\perp 1}} \frac{m_2}{2\pi k_B T_{\perp 2}}\right)^1 \iint d\boldsymbol{v}_c d\boldsymbol{v} \sigma(v) v \exp\left\{-\frac{1}{2 k_B}\left[\left(\frac{m_1}{T_{\perp 1}}\right.\right.\right.$$
$$\left.\left.\left. + \frac{m_2}{T_{\perp 2}}\right) v_{c\perp}^2 + \frac{m_r}{T_{\perp r}} v_\perp^2 + \left(\frac{m_1}{T_{\|1}} + \frac{m_2}{T_{\|2}}\right) v_{c\|}^2 + \frac{m_r}{T_{\|r}} v_{d0}^2 + \frac{m_r}{T_{\|r}} v_\|^2 + 2 \frac{m_r}{T_{\|r}} v_\| v_{d0}\right]\right\}$$

$$= \left(\frac{m_r}{2\pi k_B}\right)^{\frac{3}{2}} \frac{1}{T_{\|r}^{1/2} T_{\perp r}} \int d\boldsymbol{v} \sigma(v) v \exp\left\{-\frac{1}{2 k_B}\left[\frac{m_r}{T_{\|r}} v_\|^2 + \frac{m_r}{T_{\perp r}} v_\perp^2 + \frac{m_r}{T_{\|r}} v_{d0}^2\right.\right.$$
$$\left.\left. + 2 \frac{m_r}{T_{\|r}} v_\| v_{d0}\right]\right\}.$$

Here we have used

$$\frac{m_1 m_2}{T_1 T_2} = \left(\frac{m_1}{T_1} + \frac{m_2}{T_2}\right) \frac{m_r}{T_r}.$$

In the above, the 6D integral is reduced to 3D integral.

Further

$$E = E_\perp + E_\|, E_\| = \frac{1}{2} m_r v_\|^2, E_\perp = \frac{1}{2} m_r v_\perp^2,$$

$$v_x = \sqrt{2 E_\perp/m_r} \cos\phi, v_y = \sqrt{2 E_\perp/m_r} \sin\phi, v_z = \pm\sqrt{2 E_\|/m_r}.$$
$$\phi = [0, 2\pi], E_\perp = [0, \infty), E_\| = [0, \infty).$$

$$\begin{vmatrix} 0 & 0 & \frac{1}{2}\sqrt{\frac{2}{E_\| m_r}} \\ \frac{1}{2}\sqrt{\frac{2}{E_\perp m_r}} \cos\phi & \frac{1}{2}\sqrt{\frac{2}{E_\perp m_r}} \sin\phi & 0 \\ -\sqrt{2 E_\perp/m_r} \sin\phi & \sqrt{2 E_\perp/m_r} \cos\phi & 0 \end{vmatrix} = \frac{1}{2 m_r}\sqrt{\frac{2}{E_\| m_r}}$$

gives

$$d\boldsymbol{v} = \frac{1}{2m_r}\sqrt{\frac{2}{E_\| m_r}} dE_\perp dE_\| d\phi.$$

$$E_d = k_B T_d = \frac{m_r v_{d0}^2}{2}, \qquad v_{d0} = \sqrt{\frac{2E_d}{m_r}}.$$

Hence

$$\langle \sigma v \rangle$$
$$= \left(\frac{m_r}{2\pi k_B}\right)^{\frac{3}{2}} \frac{1}{T_{\|r}^{1/2} T_{\perp r}} \int d\boldsymbol{v} \sigma(v) v \exp\left\{-\frac{1}{2k_B}\left[\frac{m_r}{T_{\|r}}v_\|^2 + \frac{m_r}{T_{\perp r}}v_\perp^2 + \frac{m_r}{T_{\|r}}v_{d0}^2\right.\right.$$
$$\left.\left. + 2\frac{m_r}{T_{\|r}}v_\| v_{d0}\right]\right\}$$

$$= \left(\frac{m_r}{2\pi k_B}\right)^{\frac{3}{2}} \frac{1}{T_{\|r}^{1/2} T_{\perp r}} \int_0^\infty \int_0^\infty 2\pi \frac{1}{2m_r} \sqrt{\frac{2}{E_\| m_r}} dE_\perp dE_\| \, \sigma(v) \sqrt{\frac{2E}{m_r}} \left[\exp\left\{-\frac{E_\|}{k_B T_{\|r}}\right.\right.$$
$$\left.\left. -\frac{E_\perp}{k_B T_{\perp r}} - \frac{E_d}{k_B T_{\|r}} - 2\frac{\sqrt{E_\| E_d}}{k_B T_{\|r}}\right\} + \exp\left\{-\frac{E_\|}{k_B T_{\|r}} - \frac{E_\perp}{k_B T_{\perp r}} - \frac{E_d}{k_B T_{\|r}} + 2\frac{\sqrt{E_\| E_d}}{k_B T_{\|r}}\right\}\right]$$

$$= \left(\frac{1}{2\pi m_r k_B^3}\right)^{\frac{1}{2}} \frac{1}{T_{\|r}^{1/2} T_{\perp r}} \int_0^\infty \int_0^\infty dE_\perp dE_\| \, \sigma(E) \sqrt{\frac{E}{E_\|}} \exp\left\{-\frac{E_\|}{k_B T_{\|r}} - \frac{E_\perp}{k_B T_{\perp r}} - \frac{E_d}{k_B T_{\|r}}\right\}$$
$$\cdot \left[\exp\left(-2\frac{\sqrt{E_\| E_d}}{k_B T_{\|r}}\right) + \exp\left(2\frac{\sqrt{E_\| E_d}}{k_B T_{\|r}}\right)\right].$$

For $T_{\perp r} - T_{\|r} \geq 0$, define

$$E = E_\perp + E_\|, \qquad t^2 = \frac{E_\|}{k_B T_{\|r}} - \frac{E_\|}{k_B T_{\perp r}} = E_\| \frac{(T_{\perp r} - T_{\|r})}{k_B T_{\|r} T_{\perp r}},$$

i.e.,

$$E_\| = t^2 \frac{k_B T_{\|r} T_{\perp r}}{(T_{\perp r} - T_{\|r})}, \qquad E_\perp = E - t^2 \frac{k_B T_{\|r} T_{\perp r}}{(T_{\perp r} - T_{\|r})},$$

hence

$$dE_\| = 2t \frac{k_B T_{\|r} T_{\perp r}}{(T_{\perp r} - T_{\|r})} dt, \qquad dE_\perp = dE - 2t \frac{k_B T_{\|r} T_{\perp r}}{(T_{\perp r} - T_{\|r})} dt,$$

$$dE_\parallel dE_\perp = 2t\frac{k_B T_{\parallel r} T_{\perp r}}{(T_{\perp r} - T_{\parallel r})}dEdt.$$

We have

$$\langle \sigma v \rangle$$

$$= \left(\frac{1}{2\pi m_r k_B^3}\right)^{\frac{1}{2}} \frac{1}{T_{\|r}^{1/2} T_{\perp r}} \int_0^\infty \int_0^\infty dE_\perp dE_\| \, \sigma(E) \sqrt{\frac{E}{E_\|}} \exp\left\{-\frac{E_\|}{k_B T_{\|r}} - \frac{E_\perp}{k_B T_{\perp r}} - \frac{E_d}{k_B T_{\|r}}\right\}$$

$$\cdot \left[\exp\left(-2\frac{\sqrt{E_\| E_d}}{k_B T_{\|r}}\right) + \exp\left(2\frac{\sqrt{E_\| E_d}}{k_B T_{\|r}}\right)\right]$$

$$= \left(\frac{1}{2\pi m_r k_B^3}\right)^{\frac{1}{2}} \frac{1}{T_{\|r}^{1/2} T_{\perp r}} \int_0^\infty dE \int_0^{\sqrt{E\frac{(T_{\perp r} - T_{\|r})}{k_B T_{\|r} T_{\perp r}}}} dt \, \sigma(E) \sqrt{\frac{E}{t^2 \frac{k_B T_{\|r} T_{\perp r}}{(T_{\perp r} - T_{\|r})}}} \exp\left\{-\frac{t^2 \frac{k_B T_{\|r} T_{\perp r}}{(T_{\perp r} - T_{\|r})}}{k_B T_{\|r}}\right.$$

$$\left. - \frac{(E - t^2 \frac{k_B T_{\|r} T_{\perp r}}{(T_{\perp r} - T_{\|r})})}{k_B T_{\perp r}} - \frac{E_d}{k_B T_{\|r}}\right\}$$

$$\cdot \left[\exp\left(-2\frac{\sqrt{t^2 \frac{k_B T_{\|r} T_{\perp r}}{(T_{\perp r} - T_{\|r})} E_d}}{k_B T_{\|r}}\right) + \exp\left(2\frac{\sqrt{t^2 \frac{k_B T_{\|r} T_{\perp r}}{(T_{\perp r} - T_{\|r})} E_d}}{k_B T_{\|r}}\right)\right] 2t \frac{k_B T_{\|r} T_{\perp r}}{(T_{\perp r} - T_{\|r})}$$

$$= 2\left(\frac{1}{2\pi m_r k_B^2 T_{\perp r}(T_{\perp r} - T_{\|r})}\right)^{\frac{1}{2}} \int_0^\infty dE \int_0^{\sqrt{E\frac{(T_{\perp r} - T_{\|r})}{k_B T_{\|r} T_{\perp r}}}} dt \, \sigma(E) \sqrt{E} \exp\left\{-t^2 - \frac{E}{k_B T_{\perp r}}\right.$$

$$\left. - \frac{E_d}{k_B T_{\|r}}\right\} \cdot \left[\exp\left(-2t\sqrt{\frac{T_{\perp r} E_d}{k_B T_{\|r}(T_{\perp r} - T_{\|r})}}\right) + \exp\left(2t\sqrt{\frac{T_{\perp r} E_d}{k_B T_{\|r}(T_{\perp r} - T_{\|r})}}\right)\right]$$

$$= 2\left(\frac{1}{2\pi m_r k_B^2 T_{\perp r}(T_{\perp r} - T_{\|r})}\right)^{\frac{1}{2}} \int_0^\infty dE \int_0^{\sqrt{E\frac{(T_{\perp r} - T_{\|r})}{k_B T_{\|r} T_{\perp r}}}} dt \, \sigma(E) \sqrt{E} \exp\left\{-\frac{E}{k_B T_{\perp r}}\right.$$

$$\left. - \frac{E_d}{k_B T_{\|r}}\right\} \exp\left(\frac{T_{\perp r} E_d}{k_B T_{\|r}(T_{\perp r} - T_{\|r})}\right)$$

$$\cdot \left[\exp\left(-\left(t - \sqrt{\frac{T_{\perp r} E_d}{k_B T_{\|r}(T_{\perp r} - T_{\|r})}}\right)^2\right) + \exp\left(-\left(t + \sqrt{\frac{T_{\perp r} E_d}{k_B T_{\|r}(T_{\perp r} - T_{\|r})}}\right)^2\right)\right]$$

$$= \left(\frac{1}{2m_r k_B^2 T_{\perp r}(T_{\perp r} - T_{\|r})}\right)^{\frac{1}{2}} \exp\left(\frac{T_{\perp r} E_d}{k_B T_{\|r}(T_{\perp r} - T_{\|r})}\right) \int_0^\infty dE\, \sigma(E)\sqrt{E}\exp\left\{-\frac{E}{k_B T_{\perp r}} - \frac{E_d}{k_B T_{\|r}}\right\}$$

$$\cdot \left[\operatorname{erf}\left(\sqrt{E\frac{(T_{\perp r} - T_{\|r})}{k_B T_{\|r} T_{\perp r}}} + \sqrt{\frac{T_{\perp r} E_d}{k_B T_{\|r}(T_{\perp r} - T_{\|r})}}\right)\right.$$

$$\left. + \operatorname{erf}\left(\sqrt{E\frac{(T_{\perp r} - T_{\|r})}{k_B T_{\|r} T_{\perp r}}} - \sqrt{\frac{T_{\perp r} E_d}{k_B T_{\|r}(T_{\perp r} - T_{\|r})}}\right)\right]$$

$$= \left(\frac{1}{2m_r k_B^2 T_{\perp r}(T_{\perp r} - T_{\|r})}\right)^{\frac{1}{2}} \exp\left(\frac{E_d}{k_B(T_{\perp r} - T_{\|r})}\right) \int_0^\infty dE\, \sigma(E)\sqrt{E}\exp\left(-\frac{E}{k_B T_{\perp r}}\right)$$

$$\cdot \left[\operatorname{erf}\left(\sqrt{E\frac{(T_{\perp r} - T_{\|r})}{k_B T_{\|r} T_{\perp r}}} + \sqrt{\frac{T_{\perp r} E_d}{k_B T_{\|r}(T_{\perp r} - T_{\|r})}}\right)\right.$$

$$\left. + \operatorname{erf}\left(\sqrt{E\frac{(T_{\perp r} - T_{\|r})}{k_B T_{\|r} T_{\perp r}}} - \sqrt{\frac{T_{\perp r} E_d}{k_B T_{\|r}(T_{\perp r} - T_{\|r})}}\right)\right].$$

Here we have used

$$\int_{\sqrt{\frac{T_{\perp r}E_d}{k_B T_{\|r}(T_{\perp r}-T_{\|r})}}}^{\sqrt{E\frac{(T_{\perp r}-T_{\|r})}{k_B T_{\|r}T_{\perp r}}}+\sqrt{\frac{T_{\perp r}E_d}{k_B T_{\|r}(T_{\perp r}-T_{\|r})}}} \exp(-t^2)\,dt$$

$$+ \int_{-\sqrt{\frac{T_{\perp r}E_d}{k_B T_{\|r}(T_{\perp r}-T_{\|r})}}}^{\sqrt{E\frac{(T_{\perp r}-T_{\|r})}{k_B T_{\|r}T_{\perp r}}}-\sqrt{\frac{T_{\perp r}E_d}{k_B T_{\|r}(T_{\perp r}-T_{\|r})}}} \exp(-t^2)\,dt$$

$$= \frac{\sqrt{\pi}}{2}\left[\operatorname{erf}\left(\sqrt{E\frac{(T_{\perp r}-T_{\|r})}{k_B T_{\|r}T_{\perp r}}}+\sqrt{\frac{T_{\perp r}E_d}{k_B T_{\|r}(T_{\perp r}-T_{\|r})}}\right)\right.$$

$$\left. + \operatorname{erf}\left(\sqrt{E\frac{(T_{\perp r}-T_{\|r})}{k_B T_{\|r}T_{\perp r}}}-\sqrt{\frac{T_{\perp r}E_d}{k_B T_{\|r}(T_{\perp r}-T_{\|r})}}\right)\right].$$

Similarly, the above final form also holds for $T_{\|r} - T_{\perp r} > 0$.

Note

$$\operatorname{erf}(x) = 1 + \frac{e^{-x^2}}{\sqrt{\pi}}\left(-\frac{1}{x}+\frac{1}{2x^3}-\frac{3}{4x^5}+\cdots\right), x \to +\infty.$$

For $(T_{\perp r} - T_{\|r}) \to 0$

$\langle \sigma v \rangle$

$$= \left(\frac{1}{2m_r k_B^2 T_{\perp r}(T_{\perp r} - T_{\parallel r})}\right)^{\frac{1}{2}} \exp\left(\frac{E_d}{k_B(T_{\perp r} - T_{\parallel r})}\right) \int_0^\infty dE\, \sigma(E)\sqrt{E}\, \exp\left(-\frac{E}{k_B T_{\perp r}}\right)$$

$$\cdot \left[\operatorname{erf}\left(\sqrt{E\frac{(T_{\perp r} - T_{\parallel r})}{k_B T_{\parallel r} T_{\perp r}}} + \sqrt{\frac{T_{\perp r} E_d}{k_B T_{\parallel r}(T_{\perp r} - T_{\parallel r})}}\right)\right.$$

$$\left. + \operatorname{erf}\left(\sqrt{E\frac{(T_{\perp r} - T_{\parallel r})}{k_B T_{\parallel r} T_{\perp r}}} - \sqrt{\frac{T_{\perp r} E_d}{k_B T_{\parallel r}(T_{\perp r} - T_{\parallel r})}}\right)\right]$$

$$\simeq \left(\frac{1}{2m_r k_B^2 T_{\perp r}(T_{\perp r} - T_{\parallel r})}\right)^{\frac{1}{2}} \exp\left(\frac{E_d}{k_B(T_{\perp r} - T_{\parallel r})}\right) \int_0^\infty dE\, \sigma(E)\sqrt{E}\, \exp\left(-\frac{E}{k_B T_{\perp r}}\right)$$

$$\cdot \left[ -\frac{1}{\sqrt{\pi}} \left( \frac{e^{-\left(\sqrt{E\frac{(T_{\perp r}-T_{\parallel r})}{k_B T_{\parallel r} T_{\perp r}}} + \sqrt{\frac{T_{\perp r} E_d}{k_B T_{\parallel r}(T_{\perp r}-T_{\parallel r})}}\right)^2}}{\sqrt{E\frac{(T_{\perp r} - T_{\parallel r})}{k_B T_{\parallel r} T_{\perp r}}} + \sqrt{\frac{T_{\perp r} E_d}{k_B T_{\parallel r}(T_{\perp r} - T_{\parallel r})}}} \right.\right.$$

$$\left.\left. + \frac{e^{-\left(\sqrt{E\frac{(T_{\perp r}-T_{\parallel r})}{k_B T_{\parallel r} T_{\perp r}}} - \sqrt{\frac{T_{\perp r} E_d}{k_B T_{\parallel r}(T_{\perp r}-T_{\parallel r})}}\right)^2}}{\sqrt{E\frac{(T_{\perp r} - T_{\parallel r})}{k_B T_{\parallel r} T_{\perp r}}} - \sqrt{\frac{T_{\perp r} E_d}{k_B T_{\parallel r}(T_{\perp r} - T_{\parallel r})}}} \right)\right]$$

$$\simeq \left(\frac{1}{2m_r k_B^2 T_{\perp r}(T_{\perp r} - T_{\parallel r})}\right)^{\frac{1}{2}} \exp\left(\frac{E_d}{k_B(T_{\perp r} - T_{\parallel r})}\right) \int_0^\infty dE\, \sigma(E)\sqrt{E}\, \exp\left(-\frac{E}{k_B T_{\perp r}}\right)$$

$$\cdot \left[ \frac{1}{\sqrt{\frac{T_{\perp r} E_d}{k_B T_{\parallel r}(T_{\perp r} - T_{\parallel r})}}\sqrt{\pi}} e^{-\left(E\frac{(T_{\perp r}-T_{\parallel r})}{k_B T_{\parallel r} T_{\perp r}} + \frac{T_{\perp r} E_d}{k_B T_{\parallel r}(T_{\perp r}-T_{\parallel r})}\right)} \left(e^{2\frac{\sqrt{EE_d}}{k_B T_{\parallel r}}} - e^{-2\frac{\sqrt{EE_d}}{k_B T_{\parallel r}}}\right)\right]$$

$$\simeq \left(\frac{1}{2\pi m_r k_B^2 T_{\perp r} \frac{T_{\perp r} E_d}{k_B T_{\parallel r}}}\right)^{\frac{1}{2}} \int_0^\infty dE\, \sigma(E)\sqrt{E}\, \exp\left(-\frac{E}{k_B T_{\perp r}}\right)$$

$$\cdot \left[ e^{-\left(E\frac{(T_{\perp r}-T_{\|r})}{k_B T_{\|r} T_{\perp r}} + \frac{E_d}{k_B T_{\|r}}\right)} \left( e^{2\frac{\sqrt{EE_d}}{k_B T_{\|r}}} - e^{-2\frac{\sqrt{EE_d}}{k_B T_{\|r}}} \right) \right]$$

$$\simeq \left(\frac{1}{2\pi m_r k_B T_r E_d}\right)^{\frac{1}{2}} \int_0^\infty dE\, \sigma(E)\sqrt{E} \exp\left(-\frac{E}{k_B T_r} - \frac{E_d}{k_B T_r}\right)$$

$$\cdot \left( e^{2\frac{\sqrt{EE_d}}{k_B T_r}} - e^{-2\frac{\sqrt{EE_d}}{k_B T_r}} \right).$$

Hence, the result can reduce to the isotropic drift Maxwellian case.

The drift bi-Maxwellian result can unify Morse2018, Miley75, Ou2015, and Kolmes21, to a single form.

### Appendix: Calculate the kinetic energy

Kinetic energy

$$E_k = \frac{1}{2} m \int v^2 d\boldsymbol{v}\, f_j(\boldsymbol{v}_j).$$

Note

$$\int_\infty^\infty \exp\left[-\frac{(v-v_d)^2}{a^2}\right] dv = a\sqrt{\pi},$$

$$\int_\infty^\infty v^2 \exp\left[-\frac{(v-v_d)^2}{a^2}\right] dv = \frac{\sqrt{\pi}}{2} a(2v_d^2 + a^2).$$

For drift bi-Maxwellian distributions

$$f_j(v_j) = \left(\frac{m_j}{2\pi k_B T_{\|j}}\right)^{\frac{1}{2}} \left(\frac{m_j}{2\pi k_B T_{\perp j}}\right)^1 \exp\left(-\frac{m_j v_{\perp j}^2}{2k_B T_{\perp j}}\right) \exp\left(-\frac{m_j(v_{\|j}-v_{dzj})^2}{2k_B T_{\|j}}\right),$$

$$\int d\boldsymbol{v}\, f_j(\boldsymbol{v}_j) = 1,$$

gives

$E_k$

$$= \frac{1}{2} m \left(\frac{m}{2\pi k_B T_\parallel}\right)^{\frac{1}{2}} \left(\frac{m}{2\pi k_B T_\perp}\right)^{1} \iiint (v_x^2 + v_y^2 + v_z^2) \exp\left(-\frac{mv_x^2}{2k_B T_\perp}\right) \exp\left(-\frac{mv_y^2}{2k_B T_\perp}\right) \exp\left(-\frac{m(v_z - v_d)^2}{2k_B T_\parallel}\right) dv_x dv_y dv_z$$

$$= \frac{k_B T_\perp}{2} + \frac{k_B T_\perp}{2} + \frac{1}{2}(mv_d^2 + k_B T_\parallel).$$



%\bibliographystyle{apsrev4-1}
%\bibliography{sgmv_bib}

\end{document}